\documentstyle[11pt,aaspp4]{article}

\newcommand{\tastar}{\mbox{$T_A^*$}}
\newcommand{\tk}{\mbox{$T_K$}}
\newcommand{\iuv}{\mbox{$I_{UV}$}}
\newcommand{\ici}{\mbox{$\int T_A^*(C~I) dV$}}
\newcommand{\ils}{\mbox{$\int T_A^* dV$}}
\def \C{C$^{\rm o}$}
\def\gtorder{\mathrel{\raise.3ex\hbox{$>$}\mkern-14mu
\lower0.6ex\hbox{$\sim$}}}
\def\ltorder{\mathrel{\raise.3ex\hbox{$<$}\mkern-14mu
\lower0.6ex\hbox{$\sim$}}}

\begin{document}

\title{
ATOMIC CARBON IS A TEMPERATURE PROBE\\
IN DARK CLOUDS}

\author{
\sc Ken'ichi Tatematsu,\altaffilmark{1, 2}
D. T. Jaffe,
Ren\'{e} Plume,\altaffilmark{3}\\
Neal J. Evans II
}
\affil{
 Department of Astronomy, University of Texas, Austin,
TX 78712
}

\and

\author{\sc Jocelyn Keene}
\affil{
 California Institute of Technology,
320-47, Pasadena, CA 91125}

\vspace{.2in}

\altaffiltext{1}{ Present Address: Nobeyama Radio Observatory,
Nobeyama, Minamisaku, Nagano 384-1305, Japan; tatematsu@nro.nao.ac.jp}
\altaffiltext{2}{ Institute of Astrophysics and Planetary Science,
Faculty of Science, Ibaraki University,
Mito 310-0056, Japan}
\altaffiltext{3}{ Present Address:
Harvard-Smithsonian Center for Astrophysics,
60 Garden St., Cambridge, MA 02138
}

\begin{abstract}

We have mapped  the C~I $^3P_1\rightarrow$$^3P_0$ line at 492 GHz
in three molecular clouds immersed in weak ultraviolet
radiation fields, TMC-1, L134N, and IC 5146.
In all three clouds, the C~I peak $\tastar \sim 1$ K,
with very small dispersion.  
The spatial C~I distribution is extended and rather smooth.
The $J$ = 2$\rightarrow$1 transitions of CO isotopomers
were observed at the same angular resolution as C~I.
The C~I peak \tastar\  is typically a third of
the peak \tastar\  of $^{13}$CO $J$ = 2$\rightarrow$1,
and the C~I emission is usually more extended than emission in
$^{13}$CO or C$^{18}$O J=2$\rightarrow$1.
The C~I linewidth is
close to the $^{13}$CO $J$ = 2$\rightarrow$1 linewidth,
larger than the C$^{18}$O $J$ = 2$\rightarrow$1 line width,
and smaller than the $^{12}$CO $J$ = 2$\rightarrow$1 linewidth.
The shapes of these lines occasionally differ significantly,
probably because of the combined effects of differing
opacities and the physical separation of the line forming regions.
The uniformity of the C~I peak \tastar\ is remarkable for a line in 
the Wien portion of the Planck function and  indicates a very uniform 
excitation temperature. This
uniformity is best explained if the line if opaque and thermalized.
If so, the CI line probes kinetic temperature in clouds exposed to 
low ultraviolet fluxes. This conclusion has significant
implications for the thermal balance in such clouds.
At A$_V\simeq$2, these clouds have a remarkably constant
temperature from place to place and from cloud to cloud ($7.9\pm0.8$ K).  
Photodissociation region models of clouds immersed in the mean
interstellar radiation field tend to predict stronger lines than
we see, but this may be an artifact of assumptions about the temperature.
\end{abstract}

\keywords{ISM: abundances --- ISM: atoms 
--- ISM: individual (TMC-1, L134N, L183, IC 5146)
--- ISM: molecules}

\section{INTRODUCTION}

Neutral atomic carbon (\C) plays an important role in
interstellar physics and chemistry.
\C\ was recognized long ago as a potentially
important coolant for the interstellar medium
(Werner 1970; Penston 1970).
Furthermore, the abundance of \C\ may reflect
the physical structure  of molecular clouds (e.g.
Stutzki et al. 1988).

Most observations of 
the submillimeter lines
of atomic carbon
(C~I)
have 
been carried out
toward H\,{\sc ii} regions or reflection nebulae,
i.e., toward giant molecular clouds with nearby O and B stars.
Such regions, 
where the ultraviolet (UV) radiation is strong, 
are good targets to test PDR
(``photodissociation region'' or
``photon-dominated region'')
models.
Recent PDR models can explain the amount of \C\
relative to CO rather successfully
(van Dishoeck \& Black 1988;
Hollenbach, Takahashi, \& Tielens 1991).
PDR models for  uniform clouds predict that
\C\ exists in the surface layer sandwiched
between layers where C$^+$ and CO are the dominant carriers of gas-phase
carbon
(see Keene et al. 1997 for a review). 
Observationally, however, the C~I distribution
is often similar to the CO and $^{13}$CO
distribution.
Most steady-state models predict a very low abundance for
neutral carbon in the interior of the cloud (\C/CO$<$0.04 for A$_V>$2.5--4,
Jansen et al. 1995, Sternberg and Dalgarno 1995), but
Le Bourlot et al. (1993) have argued that a second solution to the
chemical equilibrium equations exists at low densities in which the
neutral carbon abundance is substantially higher (\C/CO=0.1--0.3).
Because of strong emission from the PDR, these models cannot be tested in dense,
strongly irradiated molecular cores.  In principle, one might test them
in clouds with low UV fluxes.
If substantial \C\ exists in the interiors, it may contribute to
the formation of carbon chain molecules, which are abundant in some dark clouds
(Suzuki et al. 1992).

Our goals in this work are to determine how \C\ is distributed in clouds
exposed to low UV fluxes, and to assess the contribution of such
clouds to C~I emission from galaxies.
Molecular clouds which are not
illuminated by strong UV sources
have been observed
by several researchers
(Keene et al. 1987; Stark \& van Dishoeck 1994;
Ingalls et al. 1994, 1997; Schilke et al. 1995; Stark et al. 1996).
In these studies,
fewer than ten positions toward each were observed with
small (10--15$\arcsec$) beams.
The maps of C~I emission almost all cover
scales of less than 1\arcmin\ toward objects with a size of 
hundreds of square arc minutes.

In this paper, we present more extensive C~I maps and maps of emission 
from various
CO isotopomers toward three clouds immersed in weak UV fields.
TMC1 and L134N contain no internal sources of UV radiation. 
The third source, a molecular cloud associated with IC 5146, contains an HII 
region, but the portion of the cloud we observed was
1\arcdeg\--1\arcdeg\.5 (20--30 pc) east of the exciting stars.
One can parametrize the intensity of UV radiation
field  incident on the clouds in terms of the average field
in the solar neighborhood, I$_{UV}$ (Draine 1978).
By comparing the 100 $\mu$m intensity observed toward TMC1 and L134N to
the intensity predicted from models of opaque clouds exposed to 
a UV field with a strength comparable to the field in the solar neighborhood
(Boulanger \& P\'{e}rault 1988;
Bernard et al. 1990;
Hollenbach et al. 1991), we infer that I$_{UV}\sim1$ for these two clouds.
A similar analysis for the region we mapped in IC 5146 leads to the
conclusion that I$_{UV}\sim$30.  By studying these sources with 
weak incident UV fields, we hope to increase our understanding of
how UV radiation affects the chemistry and energetics of
molecular clouds with low column densities or of the outer portions
of more opaque molecular structures.

\section{OBSERVATIONS}

Observations were carried out with the 10.4 m telescope 
of the Caltech Submillimeter Observatory (CSO)$^3$
between 1994 January and 1994 April, and in 1994 June, 1995 June, and 1998
March.
{\vskip\dimen2 \hrule \vskip\dimen1
\noindent $^3$
The CSO is operated by the California Institute of Technology
under funding from the National Science Foundation, contract AST 96-15025.
\vskip\dimen1 \hrule \vskip\dimen2
\noindent}
We observed the ground-state fine structure line of neutral carbon 
($^3P_1\rightarrow$$^3P_0$, $\nu$ = 492.1607 GHz,
Frerking et al. 1989)
as well as the $J$ = 2$\rightarrow$1
transitions of $^{12}$CO, $^{13}$CO, C$^{18}$O, and C$^{17}$O
at $219-230$ GHz,
and the $J$ = 3$\rightarrow$2 transitions of
$^{12}$CO and C$^{18}$O at $329-346$ GHz.
For simplicity, we call these two 
latter frequency bands 230 and 345 GHz.
To make C~I and CO observations of large areas on the sky 
in a limited time,
we employed a re-imaging device
for the Caltech Submillimeter
Observatory (CSO),
with which the 492 GHz receiver illuminates
only one panel of the main dish
(Plume \& Jaffe 1995; see also Plume et al. 1994, 1999 for studies of
molecular clouds with local UV sources with this instrument).
With the re-imaging device, the
measured telescope beam size ($\theta_b$) and Moon efficiency ($\eta_M$) 
are the same
at both 492 and 230 GHz: $\theta_b =  2\farcm5$ and $\eta_M = 0.8$.
At 345 GHz, these parameters have not been measured
but are assumed to be the same as those at 492 and 230 GHz.
Since the emission is very widespread and smooth, $\eta_M$ is
the appropriate efficiency.
Some positions toward L134N were also observed in the
on-the-fly (OTF) mapping mode (without
the re-imaging device), and then the data were convolved
to match the resolution of the re-imaging device.
Because the results of the OTF observations are consistent
with those of the observations with the re-imaging device,
the obtained spectra are co-added to improve the S/N ratio.
For selected positions toward TMC-1, observations without
the re-imaging device were also carried out
on 1994 March 3 and 1994 March 7.
For this work, the beam size was 15$\arcsec$ and the main-beam
efficiency
was 0.53 at 492 GHz, and the beam size was 32$\arcsec$ and the
main-beam
efficiency was 0.76 at 230 GHz.

A technical description of the 230 GHz, 345 GHz, and 492 GHz 
front-end receivers
is given by Kooi et al. (1992), Kooi et al. (1994)
and Walker et al. (1992), respectively.
The SSB system temperature was 700$-$5000 K,
700$-$1100 K, and
180$-$500 K (SSB) at 492, 345, and 230 GHz, respectively.
The receiver backend was a 1024 channel acousto-optic spectrometer. 
The total bandwidth, channel width, and 
spectral resolution were 49.7 MHz, 48.5 kHz, and 140 kHz, 
respectively.  
C~I spectra were binned to
195 kHz (0.12 km s$^{-1}$) or 390 kHz (0.24 km s$^{-1}$) channels
depending on the S/N ratio.
Data were obtained by position switching against a position shown to 
have no emission.
The spacing interval employed for mapping 
was 3$\arcmin$.
The line temperature scale was calibrated by using
the standard chopper-wheel method.  
Throughout this paper, 
we give intensities on the corrected antenna temperature
($T_A^*$) scale
(see Kutner \& Ulich 1981 
for definition).  
The calibration assumes that the signal and image 
sidebands have equal gain.  

We observed the classical ``dark'' clouds, TMC-1 and L134N (a.k.a L183) 
 and a portion of the cloud associated with IC 5146.
In addition, we use data toward
regions with local UV sources,
Orion A (this work),
Cep A, NGC 2024, S140, and W3 (Plume et al. 1999),
for comparison.
The reference center, off position, and
observed lines for each object are listed in Table 1.

\section{RESULTS}

For all the observations, we removed a first order  baseline,
determined an integrated intensity from the area under the line, and fit
a Gaussian to determine peak \tastar. While some lines were non-Gaussian,
the fits always provided a good estimate of the peak \tastar.

\subsection{TMC-1}

The distance to TMC-1 is 140 pc
(Elias 1978; Cernicharo et al. 1985).
At this distance, the beamsize of the re-imager 
($2\farcm5$) corresponds to 0.10 pc.
Figures 1(a) and (b) show the distribution of the
C~I and $^{13}$CO $J$ = 2$\rightarrow$1 spectra
on the sky.
The map center
(Table 1) is the cyanopolyyne peak 
(Churchwell, Winnewisser, \& Walmsley
1978),
and the map size is 
18$\arcmin$ (EW) $\times$ 33$\arcmin$ (NS)
(or 0.7 pc $\times$ 1.3 pc).
The peak line temperature decreases slowly away from the map center,
except to the south.
The mean  $\ici$ is 1.7$\pm$0.3 K km s$^{-1}$.
(Throughout the paper, the uncertainty reflects the standard deviation 
of the distribution, not the error in the mean.)
The corresponding values for \tastar\ are $1.1\pm0.3$ K (Fig. 12).
The ``TMC-1 ridge'', which is 
a $\sim$ 2$\arcmin$-wide NW-SE ridge passing through the
map center in high-density tracers such as NH$_3$, CCS and HC$_3$N
(Little et al. 1979; Hirahara et al. 1992),
is not prominent in C~I or $^{13}$CO.
Figure 2 shows the distribution of the \ils\  along
(a) $\Delta\alpha$ = 0$\arcsec$
and
(b) $\Delta\delta$ = 0$\arcsec$.
The distribution of \ici,
like that of $^{13}$CO $J$ = 2$\rightarrow$1,
is fairly flat across the source, while the
C$^{18}$O $J$ = 2$\rightarrow$1  and  C$^{18}$O $J$ = 1$\rightarrow$0
(Langer et al. 1995) peak more sharply at the map center.

The C~I and $^{13}$CO
\ils\ correlate very poorly in this source.
From all the points in the re-imager map of TMC-1, we obtained a 
least-squares fit to a relation between the \ils\ of:

$\ici$= (1.1$\pm 0.4) + (0.19 \pm 0.11$) $\int$ $T_A^*$($^{13}$CO) $dV$

\noindent where the correlation coefficient was 0.32.

This poor correlation contrasts with the strong correlations 
(correlation coefficient 0.85) 
found in large-scale observations of UV-illuminated molecular clouds
(Plume et al. 1999) and the even stronger correlations in the cloud
cores themselves (Keene et al. 1997).
In our large-scale TMC1 map, the C~I profile does not always resemble
the $^{13}$CO profile.
To see the difference in the line shape more clearly,
we averaged the spectra of three positions:
($\Delta\alpha$, $\Delta\delta$) =
(0$\arcsec$, $-$1260$\arcsec$),
(0$\arcsec$, $-$1080$\arcsec$), and
(0$\arcsec$, $-$900$\arcsec$).
Figure 3 shows the average spectra of
CO $J$ = 2$\rightarrow$1,
$^{13}$CO $J$ = 2$\rightarrow$1,
C$^{18}$O $J$ = 2$\rightarrow$1, and C~I.
The C~I profile is flat-topped with almost constant temperature from
5.3 to 7 km s$^{-1}$.  The $^{13}$CO J=2$\rightarrow$1 profile has a sharp 
peak at 5.5 km s$^{-1}$, a broad plateau, and then a second peak at
$\sim$6.8 km s$^{-1}$.

If some of the C~I emission arises in high column density molecular cores
and if the C$^{\rm o}$ abundance depends on the local chemistry, changes 
in that chemistry might affect the strength of the C~I emission.
To investigate this possibility,
we measured the \ils\ of C~I  and
$^{13}$CO at
the cyanopolyyne peak (0$\arcsec$, 0$\arcsec$) and at
the ammonia peak
($\Delta\alpha$, $\Delta\delta$) = ($-$290$\arcsec$, 380$\arcsec$)
(e.g., Hirahara et al. 1992; Pratap et al. 1997).
Because the ammonia peak position does not match our observation grid,
we average the spectra at ($-$360$\arcsec$, 360$\arcsec$)
and at ($-$180$\arcsec$, 360$\arcsec$).
\ici\ is 2.0 and 1.8 K km s$^{-1}$ at the cyanopolyyne and ammonia
peaks, respectively, while
the $^{13}$CO \ils\ is 
4.6 and 4.9 K km s$^{-1}$.
There is {\it no} significant peak of the C~I line at the cyanopolyyne peak; if
deep \C\ contributes to long chain carbon molecules, it is not visible
in this line.

To see the C~I line temperature and spectral variation
on smaller scales, 
we observed
the detailed line emission distribution across the TMC-1 ridge
with the 
fully illuminated CSO (beamsize 15$\arcsec$ for C~I, 
30$\arcsec$ for C$^{18}$O 
2$\rightarrow$1). 
Figure 4 shows C~I and C$^{18}$O $J$ = 2$\rightarrow$1 spectra.
The grid spacing was $\sqrt{2}~\times~20\arcsec$
from NE to SW.
The C~I and C$^{18}$O 
\ils\ distribution is rather featureless 
on this (1\arcmin) scale. The mean value of $\ici$ is 2.18  K km 
s$^{-1}$ with a standard deviation of $\pm0.36$  K km s$^{-1}$. For
C$^{18}$O J=2$\rightarrow$1, we find $\ils =1.31\pm0.19$ 
K km s$^{-1}$.
The C~I profile is broader than that of  C$^{18}$O  
$\langle \Delta$V(C~I)/$\Delta$V(C$^{18}$O)$\rangle$=1.9$\pm$0.5.
Our results are consistent with those of
Schilke et al. (1995), who observed the region with
the same telescope but with a somewhat different reference
center.

\subsection{L134N}

The distance to L134N is 
110$\pm$10 pc (Franco 1989).  The  
cloud is isolated, and it shows no
star forming activity in far-infrared observations
(Sargent et al. 1983; IRAS Point Source Catalog 1985).
L134N has been extensively observed in various molecular lines
(Swade 1989).
Phillips \& Huggins (1981) previously detected
C~I at one position and Keene (1995) has made a 20$\arcmin$ long 
strip-map of C~I in this source.
Figure 5 shows spectra obtained in an area of 
12$\arcmin$ (EW) $\times$ 30$\arcmin$ (NS)
(0.4 pc $\times$ 1.0 pc).
The beamsize 
using the re-imager corresponds to 0.08 pc.
For C~I, the $\ils$ distribution is remarkably uniform; the mean and standard
deviation are  $1.5\pm0.5$ K km s$^{-1}$.
The mean \tastar\ is $1.1\pm 0.2$ K (Fig. 12).

Figures 6(a) and (b) compare the C~I and CO isotopomer spectra
at the map center.
The C~I profile peaks at 6.8 km s$^{-1}$ and has a prominent blue wing
extending to V$_{LSR}$= 0 km s$^{-1}$. 
The $^{13}$CO, C$^{18}$O, and C$^{17}$O $J$ = 2$\rightarrow$1 and
$^{12}$CO and C$^{18}$O J=3$\rightarrow$2 lines differ significantly in
their widths, shapes, and central velocities.  These differences imply that
line opacity and/or isotopic fractionation effects influence the shapes
of most of these lines.

Figure 7 shows the \ils\ distribution
along $\Delta\alpha$ = 180$\arcsec$.
The C~I distribution is very flat, varying by only 20\%
along a strip more than half a degree (1 pc) long.
 As in TMC1, the correlation between the C~I and $^{13}$CO
\ils, based on the entire map, is poor:

$\ici$ = (0.94$\pm 0.29) + (0.14 \pm 0.07$) $\int$ $T_A^*$($^{13}$CO) $dV$

\noindent where the correlation coefficient was 0.41.

Stark et al. (1996) observed six positions
along a 10$\arcmin$-long East-West strip in L134N with the 
James Clerk Maxwell Telescope.
They found that the distribution of \ici\ is
similar to those of $^{13}$CO $J$ = 3$\rightarrow$2 and
$J$ = 2$\rightarrow$1 but not to C$^{18}$O $J$ = 2$\rightarrow$1.
Our result shows that the distribution of \ici\ is 
flatter than those of  $^{13}$CO and C$^{18}$O $J$ = 2$\rightarrow$1.
Ratios of $^{13}$CO/C~I \ils\ vary from 1.8 to 3.0 along this strip.
The  distribution of \ici\ is also substantially more extended
than the emission in the many molecular transitions observed by Swade (1989),
including C$^{18}$O $J$ =1$\rightarrow$0, CS $J$ = 2$\rightarrow$1,
NH$_3$ $J, K$ = 1, 1, H$^{13}$CO$^+$ $J$ = 1$\rightarrow$0,
and C$_3$H$_2$ $J_{K^-K^+}$ = 1$_{10}\rightarrow$1$_{01}$.

\subsection{Cases with Local Ultraviolet Sources}

\subsubsection{IC 5146}

IC 5146 (distance = 1.0 kpc, Walker 1959)
is an open cluster with an H~II region.
At 1 kpc distance, the beamsize 
of the re-imager corresponds to 0.7 pc.
A neutral cloud associated with the H~II regions has been
observed in molecular lines by
Dobashi et al. (1992), Lada et al. (1994), and Kramer et al. (1999).
Figure 8 shows spectra observed in an area of
39$\arcmin$ (EW) $\times$ 21$\arcmin$ (NS)
(11.8 pc $\times$ 6.4 pc).
The reference center is the position of IC 5146.
We have observed a region $\sim 60\arcmin - 90\arcmin$ 
away from the H II region.
Observations of this portion of IC 5146 serve as a bridge between
the study of clouds immersed in the weak ambient UV field in the 
outer Galaxy and regions illuminated by prominent local UV sources.
$\ici$ is 2.0$\pm$0.9 K km s$^{-1}$
over the mapped region. The peak T$_A^*$ is typically about 1 K
($0.9\pm0.3$, Fig. 12).
Despite our estimates of somewhat higher \iuv\ in this region, the lines
are not significantly stronger.

Figure 9 shows the \ils\ distribution
along $\Delta\delta$ = 1380$\arcsec$.
The C~I and $^{13}$CO intensities are poorly correlated.
From all the IC 5146 data, we obtain

$\ici$= (1.18$\pm 0.53) + (0.22 \pm 0.12$) $\int$ $T_A^*$($^{13}$CO) $dV$

\noindent with a correlation coefficient of 0.44.

%

\subsubsection{Other Sources}

In order to understand the differences between C~I emission from sources
with different \iuv,
we make use of large-scale C~I
and CO isotopomer maps of high-\iuv\ sources.  The data include
points from the maps of Cep A, NGC 2024, S140, and W3 made with the
re-imager by Plume et al. (1999).
In addition, as part of the present work, we have mapped a 
15$\arcmin \times 25\arcmin$ region in Orion A with the re-imaging device.
Figure 10 shows the \ils\ maps of
Orion A in
the C~I $^3P_1\rightarrow$$^3P_0$ and $^{13}$CO $J$ = 2$\rightarrow$1
emission.
\iuv\ for regions 2--15 arcmin from
$\theta^1$ C Orion ranges from 10 to 10$^5$
(Stacey et al. 1993).
The \tastar\ is much higher and more variable than that in the clouds with low
UV fluxes: mean \tastar\ is $3.8\pm1.1$.
White \& Sandell (1995) observed a smaller part of the Orion A cloud in C~I with
the James Clerk Maxwell Telescope.  With its smaller beam, they
found that the distribution of C~I is often different from that of $^{13}$CO.
Nonetheless, Keene et al. (1997) show that C~I emission correlates well
with $^{13}$CO emission in Orion at moderate intensity levels.
In our data, the C~I emission on the map is doubly peaked
around ($\Delta\alpha, \Delta\delta$) =
(0$\arcsec$, $-300\arcsec$) and (0$\arcsec$, 0$\arcsec$),
while the $^{13}$CO emission does not show double peaks.
Our map is consistent with the map of White \& Sandell
(1995),
if the difference in beamsize is taken into account.

Figure 11 plots the C~I \ils\ versus the $^{13}$CO \ils\ for the 
three dark clouds (low-\iuv) as well as for the five high-\iuv\ molecular clouds.  
A least-squares fit to the data from all of the  high-\iuv\ sources gives:

$\ici$= (3.8$\pm 0.3) + (0.25 \pm 0.02$) $\int$ $T_A^*$($^{13}$CO) $dV$

\noindent with a correlation coefficient  of 0.75, clearly a stronger correlation
than found for the low-\iuv\ clouds.  The dark cloud points cluster
among the points from the high-\iuv\ clouds with the lowest C~I
and $^{13}$CO $J$ = 2$\rightarrow$1 \ils.  The overlapping
points from the high-\iuv\ clouds mostly represent regions near
the edges of the clouds and often those parts of the clouds with the
smallest amounts of incident UV radiation (Plume et al. 1999). 
Even in aggregate, the dark clouds
show little or no correlation of C~I and $^{13}$CO \ils.

\subsection{Intensity and Linewidth}

Since the mean \tastar\ is similar in all the low-uv clouds, including
IC5146, we combined
them to find the distribution in Fig. 12; the mean \tastar\ is $1.0\pm0.3$.
This is a far lower value and a tighter distribution than is seen in 
clouds exposed to higher UV fluxes.

In contrast, the linewidths, relative to CO isotopes, are similar to those
in high-\iuv\ clouds.
Plume et al. (1999) obtained
$\langle \Delta$V(C~I)/$\Delta$V($^{13}$CO)$\rangle$ = 1.08$\pm$0.3
and
$\langle \Delta$V(C$^{18}$O)/$\Delta$V($^{13}$CO)$\rangle$ = 0.78$\pm$0.2
for massive star forming regions with local UV.
For TMC1, L134N, and IC 5146, 
we measured the linewidth ratio in the same way to be
$\langle \Delta$V(C~I)/$\Delta$V($^{13}$CO)$\rangle$ = 0.97$\pm$0.21
and
$\langle \Delta$V(C$^{18}$O)/$\Delta$V($^{13}$CO)$\rangle$ = 0.58$\pm$0.16,
which are similar to the results for sources with stronger local UV.
Here, we used spectra toward ($\Delta\alpha$, $\Delta\delta$)
= (0$\arcsec$,
$-$720$\arcsec$)$-$(0$\arcsec$, 180$\arcsec$) in TMC-1,
(0$\arcsec$, 0$\arcsec$) and (180$\arcsec$,
$-$540$\arcsec$)$-$(180$\arcsec$, 360$\arcsec$)
in L134N,
and $(-5580\arcsec, 1380\arcsec)-(-5220\arcsec, 1380\arcsec)$ 
and $(-4320\arcsec, 1380\arcsec)-(-3960\arcsec, 1380\arcsec)$
in IC 5146
where both lines were observed and the line shape is not
so much different from a Gaussian profile.
The absolute values of the linewidth are
$\Delta$V(C~I) = 1.3$\pm$0.4 km s$^{-1}$,
$\Delta$V(C$^{18}$O) = 0.7$\pm$0.2 km s$^{-1}$,
and
$\Delta$V($^{13}$CO) = 1.3$\pm$0.2 km s$^{-1}$ for TMC-1 and L134N.
Those for IC 5146 are
$\Delta$V(C~I) = 1.5$\pm$0.4 km s$^{-1}$,
$\Delta$V(C$^{18}$O) = 1.3$\pm$0.5 km s$^{-1}$,
and
$\Delta$V($^{13}$CO) = 1.8$\pm$0.4 km s$^{-1}$.
When we compare the C~I linewidth with the $^{12}$CO linewidth,
we obtain
$\langle \Delta$V(C~I)/$\Delta$V($^{12}$CO 2$\rightarrow$1)$\rangle$
 = 0.62$\pm$0.05
for ($\Delta\alpha$, $\Delta\delta$)
= (0$\arcsec$,
$-$720$\arcsec$)$-$(0$\arcsec$, 180$\arcsec$) in TMC-1.

\section{DISCUSSION}

We have detected widespread, remarkably uniform C~I emission from
 TMC1 and L134N, and from a portion of the IC 5146 cloud
far from the OB star cluster.  The C~I emission is more extended than
the emission in the $J$ = 2$\rightarrow$1 transitions of $^{13}$CO and 
C$^{18}$O and the strength of the C~I line correlates poorly with the
\ils\ for these two CO isotopomers.  In this section, we examine
the implications of these observations of clouds with low-\iuv\
for the origin of the C~I emission, for cloud chemistry, and
for the thermal structure of dark clouds.

\subsection{Photodissociation Regions Explain the Observed C~I Emission}

In our previous paper (Plume et al. 1999), we studied
the C~I $^3$P$_1\rightarrow^3$P$_0$ emission from giant molecular clouds
bathed in radiation fields 10-1000 times stronger than the mean interstellar
field in the solar neighborhood.  We showed that this
emission arises from a neutral atomic carbon layer
in a photodissociation region at the cloud surface.
The measured column densities and intensities agree well with the values
predicted by the theoretical models. Both the independence of C~I \ils\
from the strength of the UV field and the higher C~I/$^{13}$CO ratios
at the cloud edges also support the conclusion that the C~I emission comes
from the photodissociated material at the atomic/molecular interface.  
We would now like to
use the C~I and CO isotopomer maps of TMC1 and L134N, clouds with 
I$_{UV} \sim$ 1,
to examine the case for PDR's as the source for the C~I emission from
clouds immersed in lower UV fields than those incident on the GMC's.

Figure 11 plots the C~I \ils\ versus the strength of the 
$^{13}$CO J=2$\rightarrow$1 line for the positions we have observed in
TMC1 and L134N and, for comparison, the positions in IC 5146 and other clouds
illuminated by stronger local UV fields.  
The figure shows that the C~I \ils\
in TMC1 and L134N
are comparable to those observed in IC 5146, and comparable to
those at the weakest points observed
in the extended molecular clouds near OB stars observed
by Plume et al. (1999).  In this figure, we also present theoretical
C~I and $^{13}$CO \ils\ calculated on the basis of
column densities predicted by models of translucent clouds illuminated
by UV radiation (van Dishoeck and Black 1988).  
The models include the effects of isotope-selective
photodissociation and isotopic fractionation and assume a carbon depletion
factor, $\delta_C$ of 0.4.
The open squares represent 
models T3 to T6, clouds illuminated by I$_{UV}$ = 1  and 
total cloud extinction,  A$_{\rm V}^{tot}$, from 2.0 to 5.1. 
The open circles represent models I4 to I7, clouds illuminated by 
I$_{UV}$=10 
and total cloud extinction, A$_{\rm V}^{tot}$, from 2.7 to 9.0.
 We calculated the
\ils\ from the model C$^{\rm o}$ and $^{13}$CO column densities 
using a Large
Velocity Gradient (LVG)
(Scoville \& Solomon 1974; Goldreich \& Kwan 1974)
non-LTE excitation and radiative transfer code
to account, in a crude way, for excitation and opacity effects.
We used the rate coefficients for collisional excitation of C~I from
Schr\"oder et al. (1991). 
The \C\ and $^{13}$CO column densities in the models 
do not depend strongly on the kinetic temperature 
assumed by van Dishoeck and Black (see their models T6A and T6B).
Uncertainties in predicted intensities from our LVG calculations therefore
depend directly on the assumed kinetic temperature rather than indirectly
through linewidth-dependent photochemical effects.  In this rough comparison,
we have used T$_K$=10 K to be consistent with measures of core temperatures
in dark clouds (e.g. Benson \&
Myers 1980). 
We set the column density per unit velocity in the radiative transfer 
model equal to the \C\ or $^{13}$CO column density from the theoretical
model divided by the typical linewidth in the clouds.  The results
in Figure 11, where we have multiplied the model \ils\ by the
beam efficiency to compare to the observations, show that the van Dishoeck
and Black models for I$_{UV}$=1 correctly predict the C~I \ils\
we observe in the maps of TMC1 and L134N.  The observed
$^{13}$CO \ils\ are comparable to the values for the models 
with the largest A$_{\rm V}^{tot}$.

The morphology of the C~I and molecular emission from the 
 clouds with I$_{UV} \sim$ 1 also
supports the idea that the C~I line arises in material at the cloud 
surface.  In both L134N and TMC1, the C~I \ils\ distribution is
broader and flatter than that of $^{13}$CO and significantly broader than 
the C$^{18}$O J=2$\rightarrow$1 distribution.

\subsection{Why Don't C~I and Molecular Line Intensities Correlate in
Low-\iuv\ Clouds?}

If the C~I emission arises in a thin layer near the cloud surface and 
emission from molecular lines less opaque than the lowest few transitions
of $^{12}$CO arise throughout the bulk of the cloud, there is no
{\it a priori} reason to expect the C~I and $^{13}$CO or C$^{18}$O
\ils\ to correlate.  Nevertheless, the C~I and CO isotopomer
\ils\ correlate astonishingly well in high column density
cloud cores illuminated by powerful UV fields like M17 SW (Keene et
al. 1997). There is a clear difference in the \ils\ correlation
between high-\iuv\ clouds (correlation coefficient = 0.85 for C~I vs. 
$^{13}$CO \ils\, Plume et al. 1999) and
the low-\iuv\ dark clouds studied here (correlation coefficient = 0.32 and 0.41
for C~I vs. $^{13}$CO in TMC1 and L134N, respectively).

For cloud cores, Plume et al. (1999) point out the similarity of the
clump/interclump structure needed to produce the observed extent of the
C~II 158 $\mu$m line emission to the kind of structure that would
produce the high degree of spatial correlation between C~I and $^{13}$CO
seen by Keene et al. (1997); in both cases the \ils\ reflect the
number of clumps in the beam.  Plume et al. argue that in extended GMC
material, unlike the high column density cores,
 the C~I lines arise predominantly in a global layer over the 
entire face of the cloud.  They explain the somewhat weaker
C~I--$^{13}$CO \ils\ correlation 
as a conspiracy of edge effects and the correlation of temperature and
density variations with total column density.  The lack of intensity 
correlations in TMC1 and L134N, then, results from an absence of 
temperature variations as well as from higher opacities in C~I
$^3$P$_1\rightarrow^3$P$_0$ and the lowest transitions of $^{13}$CO
and C$^{18}$O at low temperature (see below).
As is so dramatically illustrated in the small-scale C~I--molecular line
comparisons toward the TMC1 ridge (Figure 4 and Schilke et al.), the
C~I observations toward clouds illuminated by I$_{UV} \sim$1 external fields
 do not contain any information 
about high column density molecular regions.

\subsection{Thermal Balance: The Meaning of Constant C~I Line Brightness}

The most striking feature of the data are the very narrow distributions
of \tastar\ seen in Fig. 12. Such tight distributions are not seen
in any molecular line besides CO, which is opaque and thermalized. We
suggest that the CI lines are also opaque and thermalized; thus the
values of \tastar\ provide a probe of \tk\ in the PDR layer with $\tau \sim 1$.

There are both observational and theoretical arguments for a high 
opacity in the C~I $^3$P$_1\rightarrow^3$P$_0$ line.  The large 
linewidth and, in TMC1, flat-topped shape of the C~I lines are both
indications that the opacity is high. Furthermore, the peak intensity
does not vary strongly despite significant variations in line width
(Figures 1a, 5, and 8).  If the lines were not thermalized, \tastar\
would be very sensitive to density fluctuations and the tight distribution
seen in Fig. 12 would be extremely unlikely.
To investigate the C~I opacity more thoroughly, we also
computed spherical Monte Carlo models (Choi et al. 1995) for 
photodissociated clouds.  The temperature and C~I abundance versus
distance output from a plane parallel PDR model with I$_{UV}$=1 and
n$_{H+2H_2}$=10$^3$ cm$^{-3}$ was mapped onto a spherical cloud
(see Pak et al. 1998, van Dishoeck \& Black 1987) with a microturbulent
velocity dispersion of 0.65 km s$^{-1}$.  The model, which has a
line center opacity $\sim$2, correctly reproduces the linewidths
seen in the $^3$P$_1\rightarrow^3$P$_0$ spectra. 
Assumptions about the cloud core temperature (see below) may mean that
the C~I opacity is even higher. Because the
critical density for the $^3$P$_1\rightarrow^3$P$_0$ transition is 
only 800 cm$^{-3}$ and the photon escape probabilities are substantially
lower than unity, it
is very likely that the transition is thermalized ($T_{ex} = T_K$).

If the C~I line is opaque and thermalized, it provides a probe of \tk\
deep in the PDR layer, where $\tau \sim 1$. The distribution of
\tk\ is shown in Fig. 13 for all the dark clouds and for Orion A.
The distribution for the dark clouds is remarkably tight:
mean \tk\ is $7.9\pm0.8$ K, a 10\% variation. The high frequency of the C~I
$^3$P$_1\rightarrow^3$P$_0$ transition means that it is in the Wien
limit for the temperatures we derive; thus a tight distribution 
in \tastar\ (30\% variation) becomes even tighter in \tk.
Small variations in temperature would lead to an enormous fractional change
in observed T$_A^*$ (T$_K$=8$\pm$3 K implies, for thermalized gas,
a T$_A^*$ range from 0.17 to 2.5 K, a factor of 15). In comparison,
the range of \tk\ inferred for Orion is much broader, consistent
with a lower and more variable opacity; thus the values of \tk\
in Orion are lower limits.

Do these values of \tk\ agree with predictions of PDR models for low-\iuv?
The PDR models discussed above produce
T$_K\sim$11.5 K in the line forming region.  This higher T$_K$ results in
a $\sim$3 K C~I line.  The high predicted temperature may be, in part,
an artifact of the assumption that the asymptotic cloud core
temperature was 10 K or of slight differences between the actual and
theoretical thermal balance. Recent models by Kaufman et al. (1999)
find that \tk\ drops below 10 K for very small ratios of \iuv\ to 
density. Comparison to their predictions of intensity finds agreement with our
observations only at low \iuv\ and quite high density ($n > 10^5$ cm$^{-3}$).
Is there other evidence for such low \tk\ in low-\iuv\ clouds?
Clemens, Yun, \& Heyer (1991) found that 74\% of a sample of such
clouds have $\tk = 8.5$ K.

Why do the C~I layers have such a constant temperature?  
Given the inevitability of density fluctuations,
what causes the thermal balance to choose this temperature?  One strong 
possibility is that neutral carbon itself provides the thermostat.
The fact that the C~I line is in the Wien portion of the Planck
function at these low \tk\ means that the cooling power in the
492 GHz line rises very rapidly with temperature in the 5-10 K range.
In the specific layer where C~I is present, the higher opacity of the
$^{12}$CO line and the low abundance of $^{13}$CO may keep the
rotational lines of CO isotopomers from contributing strongly to the
cooling.  Indeed models by Hollenbach, Takahashi, \& Tielens (1991)
indicate that C~I cooling does dominate in the layer where neutral 
atomic carbon is
abundant.

If our interpretation is correct, the
higher transition of C~I, $^3P_2\rightarrow$$^3P_1$ at 809 GHz,
will be very weak toward these clouds. An upper limit
can be obtained by assuming that the higher transition is
also opaque and thermalized. In this case,  
the radiation temperature (or $T_A^*/\eta_M$) is 0.3 K.
However, the higher line is very unlikely to be opaque.
Even if the levels are thermalized, the optical depth in 
the higher line will be only 0.16 that of the lower line.
More realistic estimates can be obtained from models of clouds in weak
UV fields, combined with LVG calculations.
Model T6 ($n$ = 10$^3$ cm$^{-3}$,
$A_V^{tot}$ = 5.1 mag) of van Dishoeck \& Black (1988)
and LVG calculation
with $\tk = 8$ K predict the
$^3P_1\rightarrow$$^3P_0$ and $^3P_2\rightarrow$$^3P_1$
radiation temperatures
to be 0.9 K and 0.02 K, respectively.
Model T6C
with higher density ($n$ = 10$^4$ cm$^{-3}$)
and LVG calculation with $\tk = 8$ K
predict the
$^3P_1\rightarrow$$^3P_0$ and $^3P_2\rightarrow$$^3P_1$
radiation temperatures to be
1.1 and 0.04 K, respectively.
The expected ratio of $^3P_2\rightarrow$$^3P_1$ to
$^3P_1\rightarrow$$^3P_0$ is thus less than 0.05. 
To the authors' knowledge, no observations
of C~I $^3P_2\rightarrow$$^3P_1$ toward molecular
clouds in weak UV fields have been reported in the literature.

\subsection{Overall Cooling by C~I}

The C~I $^3P_1\rightarrow$$^3P_0$
line was expected to play an important role in
interstellar gas cooling (Penston 1970).
The relevant quantity here is $F = \int\int I_{\nu} d\nu d\Omega$; 
because $F \propto \nu^3 \ils$, C~I may be more important than
one might guess from the ratios of $\ils$.
We derive $F$ (C~I)/$F$ ($^{12}$CO $J$ = 2$\rightarrow$1)
= 1.9$\pm$0.3 for TMC-1 and 1.0$\pm$0.5 for IC 5146.
The ratio $F$ (C~I)/$F$ ($^{12}$CO $J$ = 3$\rightarrow$2)
in L134N is 0.8$\pm$0.2.  In these dark clouds,
the cooling power of the C~I $^3P_1\rightarrow$$^3P_0$
line is as important as low-$J$ $^{12}$CO lines.
The flux ratios in the average spectrum of the Galaxy (Wright et al. 1991)
are $F$ (C~I)/$F$ ($^{12}$CO $J$ = 2$\rightarrow$1) = 2.3 and
$F$ (C~I)/$F$ ($^{12}$CO $J$ = 3$\rightarrow$2) = 1.6.
%

\section{Summary}

We have mapped the C~I $^3$P$_1\rightarrow^3$P$_0$ transition in three clouds
with low-I$_{UV}$ values at their surfaces.  The C~I emission is more
extended than emission in low-J CO isotopomer lines and the intensity
is very uniform.
C~I linewidths are significantly larger than C$^{18}$O linewidths and 
comparable to or larger than the widths of low-J $^{13}$CO lines. 
The uniform intensity and large linewidth of the C~I lines, as well as models 
of low-I$_{UV}$ photodissociation regions imply that the 
C~I $^3$P$_1\rightarrow^3$P$_0$ transition is thermalized and optically 
thick.  The uniform 
T$_A^*$=1.0$\pm$0.3 K, together with the high frequency of the C~I line,
means that the kinetic temperature in the emitting layer is remarkably
constant, T$_K$(C~I layer)=7.9$\pm$0.8 K.  We suggest that the C~I 
emission itself acts as a thermostat in controlling the temperature of 
this layer in low-I$_{UV}$ clouds.

\acknowledgments

The authors would like to thank 
Wenbin Li for his help during the observations and
the CSO staff for their technical assistance.
Thanks are also due to
Jill Knapp, Ken Young, Merc\`{e} Crosas, and Zeljko Ivezic
for letting us use a part of their telescope time.
This work was partially supported by
the NSF Grants AST-9017710 and AST-9530695 to the University of Texas
 and by the David and Lucile 
Packard Foundation.
K. T. was supported by
Grants-in-Aid from the Ministry of Education, Science,
Sports, and Culture of Japan (Nos. 07740180 and 11440067)
and by a grant from the Sumitomo Foundation (No. 950570).

\clearpage

\begin{table*}
\begin{center}
\caption{\sc Reference Center and Off Position} \label{tbl-1}
\begin{tabular}{lllrrl}
\tableline
\tableline
Source & Reference & Center &  Off & Position \\
 & $\alpha (1950)$ & $\delta (1950)$ & $\Delta\alpha$ & $\Delta\delta$ 
 & Observed Line\\
\tableline
TMC-1 & 4$^h$38$^m$38$^s$.6 & 25\arcdeg\ 36$\arcmin$ 00$\arcsec$ 
& 2400$\arcsec$ & 0$\arcsec$ & C~I,\\
 & & & & & $J$ = 2$\rightarrow$1 of $^{12}$CO, $^{13}$CO, C$^{18}$O\\
Orion A & 5$^h$32$^m$47$^s$ & $-$5\arcdeg\ 24$\arcmin$ 23$\arcsec$
& 1800$\arcsec$ & 1800$\arcsec$ & C~I,\\
 & & & & & $J$ = 2$\rightarrow$1 of $^{13}$CO, C$^{18}$O\\
L134N & 15$^h$51$^m$30$^s$ & $-$2\arcdeg\ 43$\arcmin$ 31$\arcsec$ 
& $-660\arcsec$ & 660$\arcsec$
& C~I,\\
 & & & & & $J$ = 2$\rightarrow$1 of $^{13}$CO, C$^{18}$O, C$^{17}$O,\\
 & & & & & $J$ = 3$\rightarrow$2 of C$^{18}$O\\
$\arcsec$ &$\arcsec$ &$\arcsec$ 
& $-1800\arcsec$ & 0$\arcsec$ \tablenotemark{a}
 & $J$ = 3$\rightarrow$2 of $^{12}$CO\\
IC 5146 & 21$^h$51$^m$54$^s$ & 47\arcdeg\ 01$\arcmin$ 12$\arcsec$ 
& $-3000\arcsec$ & 0$\arcsec$ & C~I,\\
 & & & & & $J$ = 2$\rightarrow$1 of $^{12}$CO, $^{13}$CO, C$^{18}$O\\
\tableline

\end{tabular}
\end{center}
\tablenotetext{a}{ Off position is in ($\Delta$AZ, $\Delta$EL) for this 
line only}
\end{table*}

\clearpage

\begin{figure}
\caption{
Line profiles toward TMC-1. The (0,0) position was 
$\alpha$(1950)= 4$^h$38$^m$38$^s$.6, $\delta$(1950)= 25\arcdeg\ 36$\arcmin$ 00$\arcsec$. These data were taken with the 
re-imager with a 150$\arcsec$ beam. The intensity scale is T$_A^*$.
(a) C~I $^3P_1\rightarrow$$^3P_0$, rebinned to 0.24 km s$^{-1}$ channel
spacing.
(b) $^{13}$CO $J$ = 2$\rightarrow$1.
}
\end{figure}

\begin{figure}
\caption{
Distribution of \ils\ toward TMC-1,
along (a) $\Delta\alpha$ = 0$\arcsec$
(b) $\Delta\delta$ = 0$\arcsec$.
Open circles, open boxes, and filled circles
represent the \ils\ of $^{13}$CO $J$ = 2$\rightarrow$1,
C$^{18}$O $J$ = 2$\rightarrow$1,
and C~I $^3P_1\rightarrow$$^3P_0$, respectively.
The error bars reflect only the rms noise level.
}
\end{figure}

\begin{figure}
\caption{
Line profiles
observed with the re-imaging device toward TMC-1.
C~I spectra were binned to 0.24 km s$^{-1}$ channels,
and some of the CO isotopomer spectra were binned to 0.12 km s$^{-1}$
channels.  The spectra are sums of data taken at three positions:
($\Delta\alpha$, $\Delta\delta$) =
(0$\arcsec$, $-$1260$\arcsec$),
(0$\arcsec$, $-$1080$\arcsec$), and
(0$\arcsec$, $-$900$\arcsec$).
}
\end{figure}

\begin{figure}
\caption{
C~I $^3P_1\rightarrow$$^3P_0$ (left)
and C$^{18}$O $J$ = 2$\rightarrow$1 (right)
spectra obtained toward TMC-1
with the full CSO telescope (beamsize 15$\arcsec$ for C~I and 32$\arcsec$
for C$^{18}$O J=2$\rightarrow$1).
}
\end{figure}

\begin{figure}
\caption{
C~I line profiles toward L134N.
 These data were taken with the 
re-imaging device with a 150$\arcsec$ beam. The intensity scale is T$_A^*$.
The (0,0) position was $\alpha$(1950)= 15$^h$51$^m$30$^s$,
$\delta$(1950)= $-$2\arcdeg\ 43$\arcmin$ 31$\arcsec$ 
Spectra were binned to 0.24 km s$^{-1}$.
}
\end{figure}

\begin{figure}
\caption{
A comparison of
line profiles
observed with the re-imaging device at the map center for L134N.
(a) The $J$ = 2$\rightarrow$1 transitions of $^{13}$CO, C$^{18}$,
and C$^{17}$O, and C~I.
C~I, C$^{18}$O, and C$^{17}$O
spectra were binned to 0.24 km s$^{-1}$ channels.
(b) The $J$ =3$\rightarrow$2 transitions of $^{12}$CO and C$^{18}$O.
The spectra were binned to 0.24 km s$^{-1}$ channels.
}
\end{figure}

\begin{figure}
\caption{
Distribution of \ils\ 
along
$\Delta\alpha$ = 180$\arcsec$ in L134N.
The symbols are the same as those used in Figure 2.
}
\end{figure}

\begin{figure}
\caption{
C~I line profiles toward IC 5146.
The (0,0) position for this map was 
$\alpha$(1950)= 21$^h$51$^m$54$^s$ 
$\delta$(1950)= 47\arcdeg\ 01$\arcmin$ 12$\arcsec$.  
}
\end{figure}

\begin{figure}
\caption{
Distribution of the \ils\
along
$\Delta\delta$ =1380$\arcsec$ in IC 5146.
The symbols are the same as those used in Figure 2.
}
\end{figure}

\begin{figure}
\caption{
Map of \ils\ of C~I $^3P_1\rightarrow$$^3P_0$
(solid lines)
and $^{13}$CO $J$ 2$\rightarrow$1
(dotted lines)
toward Orion A.
The contour interval is 3 K km s$^{-1}$
for C~I 
and
10 K km s$^{-1}$ for $^{13}$CO.
}
\end{figure}


\begin{figure}
\caption{
A plot of C~I \ils\ against that 
of $^{13}$CO J=2$\rightarrow$1.
The filled diamonds, triangles, and boxes
represent values observed toward 
TMC-1, L134N, IC 5146, 
respectively.  The dots represent points from the GMC's observed
for this paper and by Plume et al. (1999).
The connected open symbols represent C~I and $^{13}$CO \ils\ predicted 
by an LVG model and column densities from the PDR chemistry models of
van Dishoeck \& Black (1988).  The open squares represent models T3-T6,
for I$_{UV}$=1  and total cloud extinction, A$_{\rm V}^{tot}$, from 2.0 to 5.1.
The open circles represent models I4-I7, for I$_{UV}$=10 
and total cloud extinction, A$_{\rm V}^{tot}$, from 2.7 to 9.0.
}
\end{figure}

\begin{figure}
\caption{
The lower three panels show the distribution of \tastar\ for the three
low-\iuv\ clouds, with the mean and standard deviation indicated.
The distribution for all three dark clouds is shown in the upper left, 
while that for Orion (note very different scale) is shown in the upper
right.
}
\end{figure}

\begin{figure}
\caption{
The distribution of \tk, assuming that the lines are opaque and thermalized,
is shown for all the dark clouds and for Orion A. The assumption that the 
line is opaque is probably not valid in Orion A, so these represent lower
limits to \tk.
}
\end{figure}


\clearpage

\begin{references}
\reference{} Benson, P. J., \& Myers, P. C. 1980, \apj, 242, L87
\reference{} Bernard, J. P., D\'{e}sert, F.-X., \& Boulanger, F. 1990,
in The Interstellar Medium in External Galaxies,
ed. D. Hollenbach \& H. Thronson (NASA CP-3084), 105
\reference{} Boulanger, F., \& P\'{e}rault, M. 1988,
\apj, 330, 964
\reference{} Cernicharo, J., Bachiller, R., \& Duvert, G. 1985, \aap, 149,
273
\reference{} Choi, M., Evans, N. J., II, Gregersen, E. M., \& Wang, Y.
1995, \apj, 448, 742
\reference{} Churchwell, E., Winnewisser, G. \& Walmsley, C. M.
1978, \aap, 67, 139
\reference{} Clemens, D. P., Yun, J. L., \& Heyer, M. H. 1991, \apjs, 75, 877
\reference{} Dobashi, K., Yonekura, Y., Mizuno, A., \& Fukui, Y.
1992, \aj, 104, 1525
\reference{} Draine, B. T. 1978, \apjs, 36, 595
\reference{} Elias, J. M. 1978, \apj, 224, 857
\reference{} Franco, G. P. 1989, \aap, 223, 313
\reference{} Frerking, M. A., Keene, J., Blake, G. A., \& Phillips, T. G.
1989, \apj, 344, 311
\reference{} Goldreich, P., \& Kwan, J. Y. 1974, \apj, 189, 441
\reference{} Hirahara, Y., et al. 1992, \apj, 394, 539
\reference{} Hollenbach, D. J., Takahashi, T., \& Tielens, A. G. G. M.
1991, \apj, 377, 192
\reference{} Ingalls, J. G., Bania, T. M., \& Jackson, J. M. 1994,
\apj, 431, L139
\reference{} Ingalls, J. G., Chamberlin, R. A., Bania, T. M.,
\& Jackson, J. M. 1997, \apj, 479, 296
\reference{} IRAS Point Source Catalog. 1985, Joint IRAS Science Working
Group, (Washington, DC: US Goverment Printing Office)
\reference{} Jansen, D.J., van Dishoeck, E.F., Black, J.H., Spaans, M., \& 
Sosin, C. 1995, \aap, 302, 223
\reference{} Kaufman, M. J., Wolfire, M. G., Hollenbach, D. J., \& Luhman, M. L.
 1999, preprint
\reference{} Keene, J. 1995, in The Physics and Chemistry of Interstellar Molecular clouds, ed. G. Winnewisser \& G.C. Pelz, (Springer: Berlin), p. 186
\reference{} Keene, J., Blake, G. A., \& Phillips, T. G. 1987,
\apj, 313, 396
\reference{} Keene, J., Lis, D. C., Phillips, T. G., \& Schilke, P.
1997, Molecules in Astrophysics: Probes and Processes,
ed. E. F. van Dishoeck (Dordrecht: Kluwer), 129
\reference{} Keene, J., Schilke, P., Kooi, J., Lis, D. C.,
Mehringer, D., \& Phillips, T. G. 1998, \apj, 494, L107
\reference{} Kooi, J. W., Chan, M., Bumble, B., Phillips, T. G.,
Schaffer, P. L., \& LeDuc, H. G.
1994, Int. J. IR and MM Waves, 15, 783
\reference{} Kooi, J. W., Chan, M., Phillips, T. G., Bumble, B., 
\& LeDuc, H. G.
1992,  IEEE Trans. Microwave Theory Tech., 40, 812
\reference{} Kramer, C., Alves, J., Lada, C. J., Lada, E. A., Sievers, A.,
Ungerechts, H., \& Walmsley, M. 1999, \aap, 342, 257
\reference{} Kutner, M. L., \& Ulich, B. L. 1981, \apj, 250, 341
\reference{} Lada, C. J., Lada, E. A., Clemens, D. P., \& Bally, J.
1994, \apj, 429, 694
\reference{} Langer, W. D., Velusamy, T., Kuiper, T. B. H.,
Levin, S., Olsen, E., \& Migenes, V. 1995,
\apj, 453, 293
\reference{} Le Bourlot, J., Pineau des For\^{e}ts, G., Roueff, E.
\& Schilke, P. 1993, \apj, 416, L87
\reference{} Little, L. T., MacDonald, G. H., Riley, P. W.,
\& Matheson, D. N. 1979, \mnras, 189, 539
\reference{} Pak, S., Jaffe, D. T., van Dishoeck, E. F.,
Johansson, L. E. B., \& Booth, R. S. 1998,
\apj, 498, 735
\reference{} Penston, M. V. 1970, \apj, 162, 771
\reference{} Phillips, T. G., \& Huggins, P. J. 1981, \apj, 251, 533
\reference{} Plume, R., \& Jaffe, D. T. 1995,
\pasp, 107, 488
\reference{} Plume, R., Jaffe, D. T., \& Keene, J. 1994, ApJ, 425, L49
\reference{} Plume, R., Jaffe, D. T., Tatematsu, K., Evans, N. J., II,
\& Keene, J.
1999, ApJ, 512, 768
\reference{} Pratap, P., Dickens, J. E., Snell, R. L., Miralles, M. P.,
Bergin, E. A., Irvine, W. M., \& Schloerb, F. P. 1997, \apj, 486, 862
\reference{} Sargent, A. I., van Duinen, R. J., Nordh, H. L.,
Fridlund, C. V. M.,
Aalders, J. W. G. \& Beintema, D. 1983, \aj, 88, 88
\reference{} Schilke, P., Keene, J., Le Bourlot, J., Pineau des
For\^{e}ts, G., \& Roueff, E. 1995, \aap, 294, L17
\reference{} Schr\"{o}der, K., Staemmler, V., Smith, M. D.,
Flower, D. R., \& Jaquet, R. 1991, J. Phys. B, 24, 2487
\reference{} Scoville, N. Z., \& Solomon, P. M. 1974, \apj, 187, L67
\reference{} Stacey, G. J., Jaffe, D. T., Geis, N., Genzel, R., Harris, A.
I.,
Poglitsch, A., Stutzki, J., \& Townes, C. H. 1993, \apj, 404, 219
\reference{} Stark, R., \& van Dishoeck, E. F. 1994, \aap, 286, L43
\reference{} Stark, R., Wesselius, P. R., van Dishoeck, E. F.,
\& Laureijs, R. J. 1996, \aap, 311, 282
\reference{} Sternberg, A. \& Dalgarno, A. 1995, \apjs, 99, 565
\reference{} Stutzki, J., Stacey, G. J., Genzel, R., Harris, A. I.,
Jaffe, D. T., \& Lugten, J. B. 1988, \apj, 332, 379
\reference{} Suzuki, H., Yamamoto, S., Ohishi, M., Kaifu, N., 
Ishikawa, S., Hirahara, Y., \& Takano, S.
1992, \apj, 392, 551
\reference{} Swade, D. A. 1989, \apjs, 71, 219
\reference{} van Dishoeck, E. F., \& Black, J. H. 1988, \apj, 334, 771
\reference{} Walker, C. K., Kooi, J. W., Chan, M., LeDuc, H. G.,
Carlstrom, J. E., \& Phillips, T. G. 1992, Int. J. IR \& MM Waves,
13, 785
\reference{} Walker, M. F. 1959, \apj, 130, 57
\reference{} Werner, M. W. 1970, Astrophys. Lett., 6, 81
\reference{} White, G. J., \& Sandell, G. 1995, \aap, 299, 179
\reference{} Wright, E. L., et al. 1991, \apj, 381, 200
\end{references}
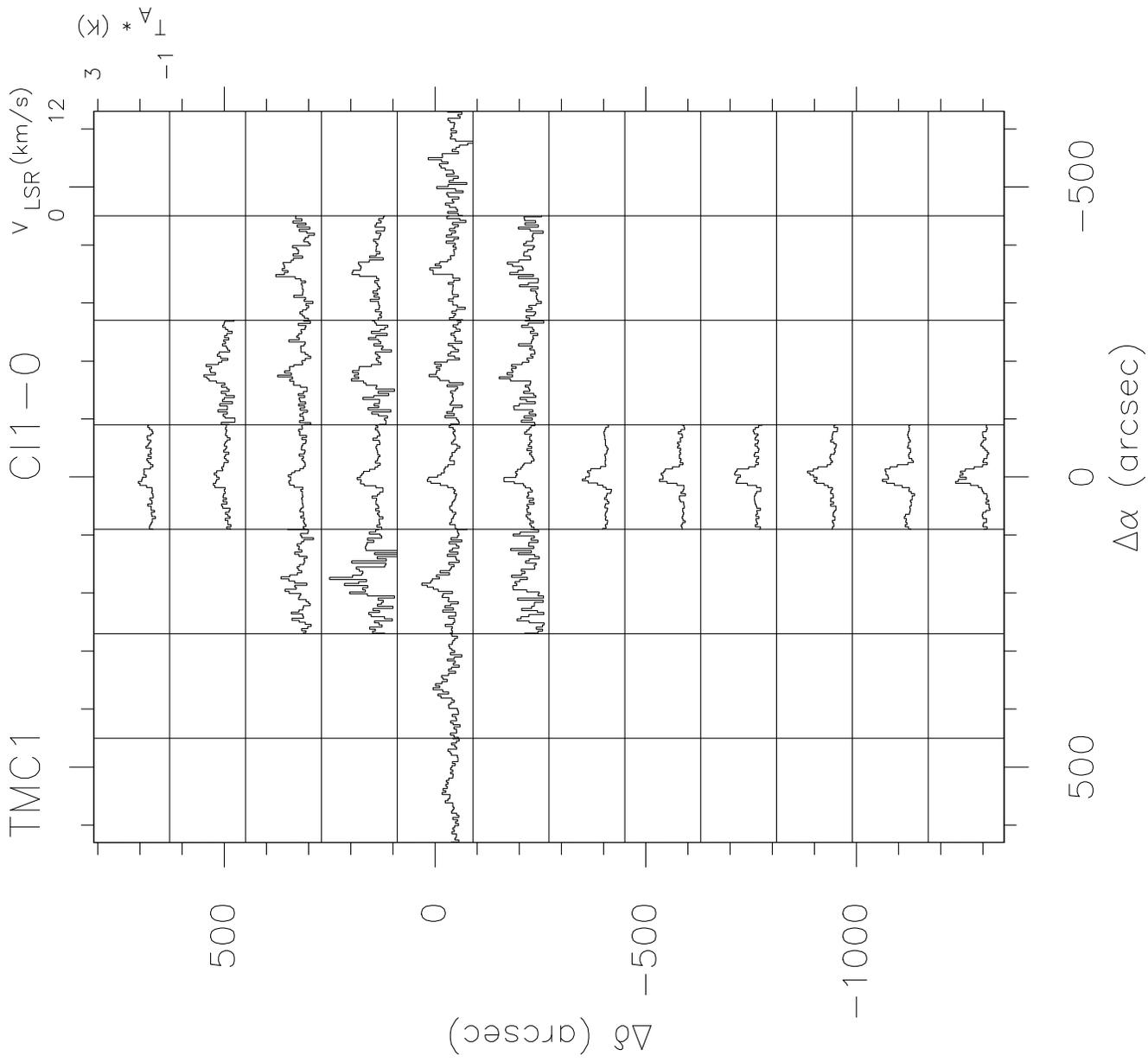
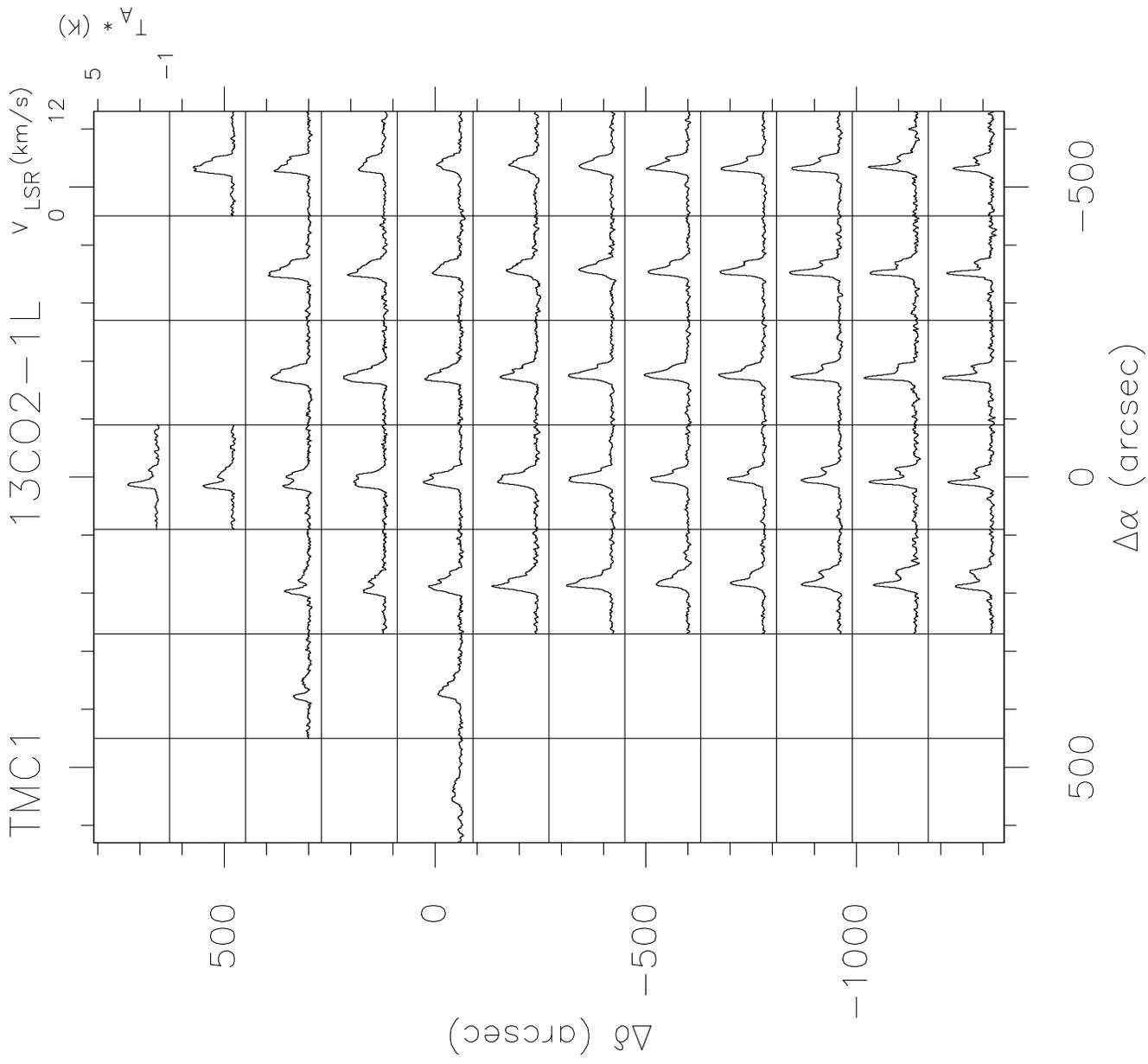
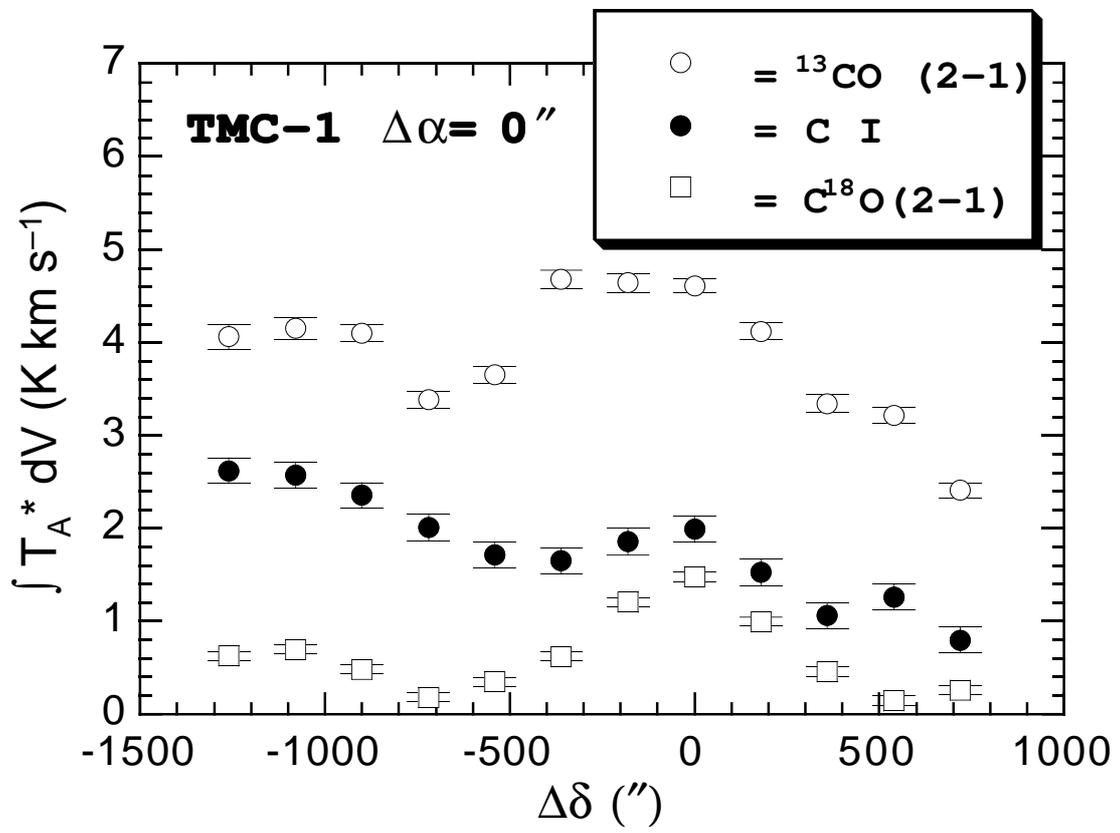
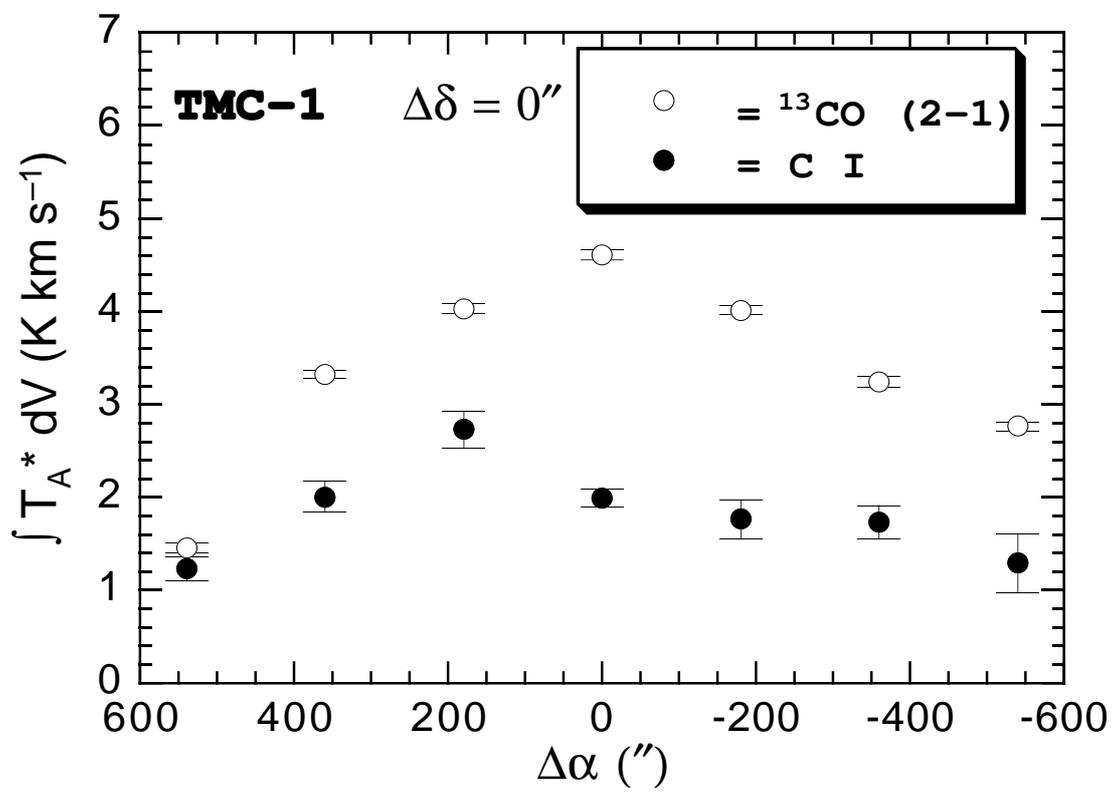
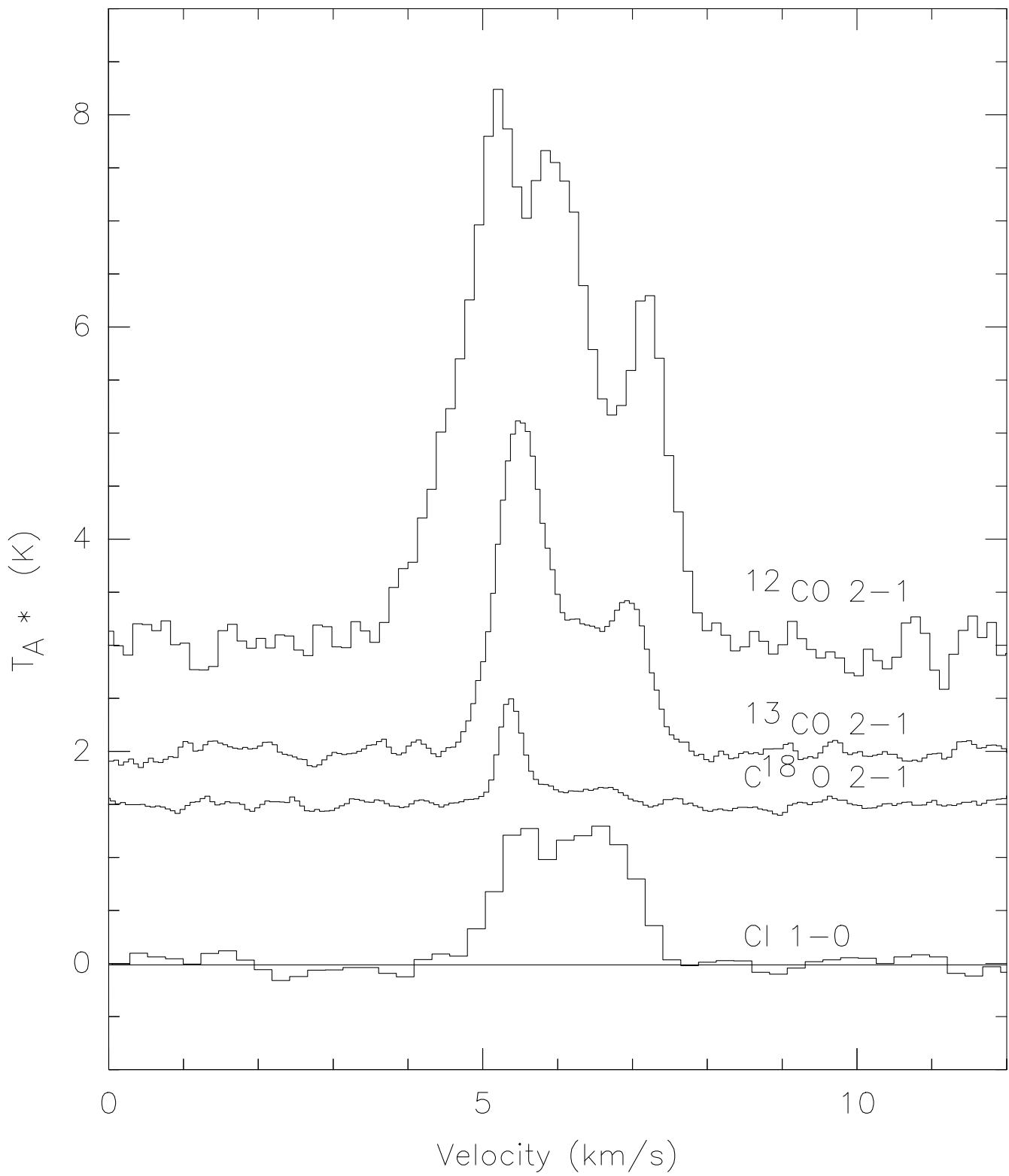
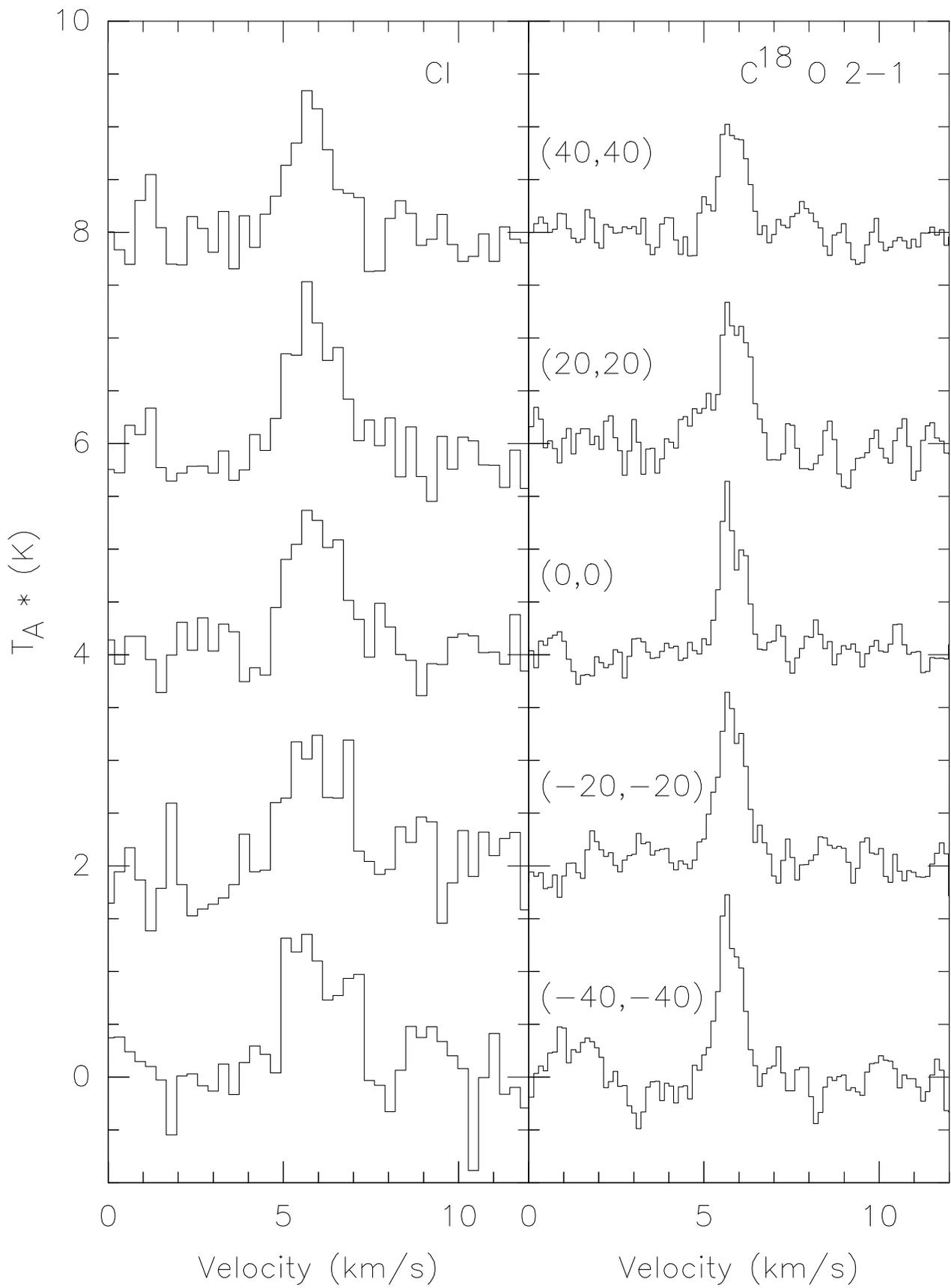
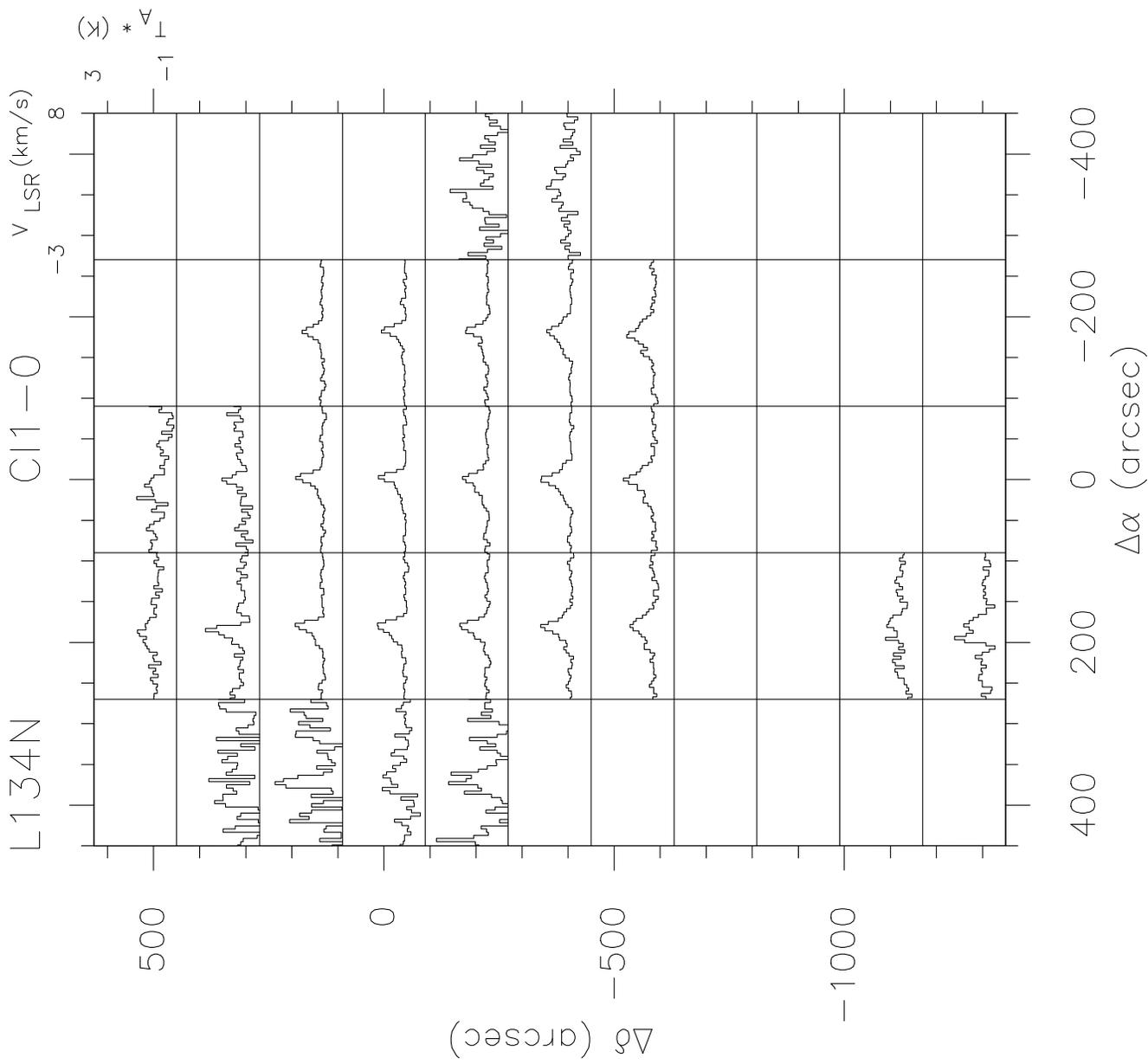
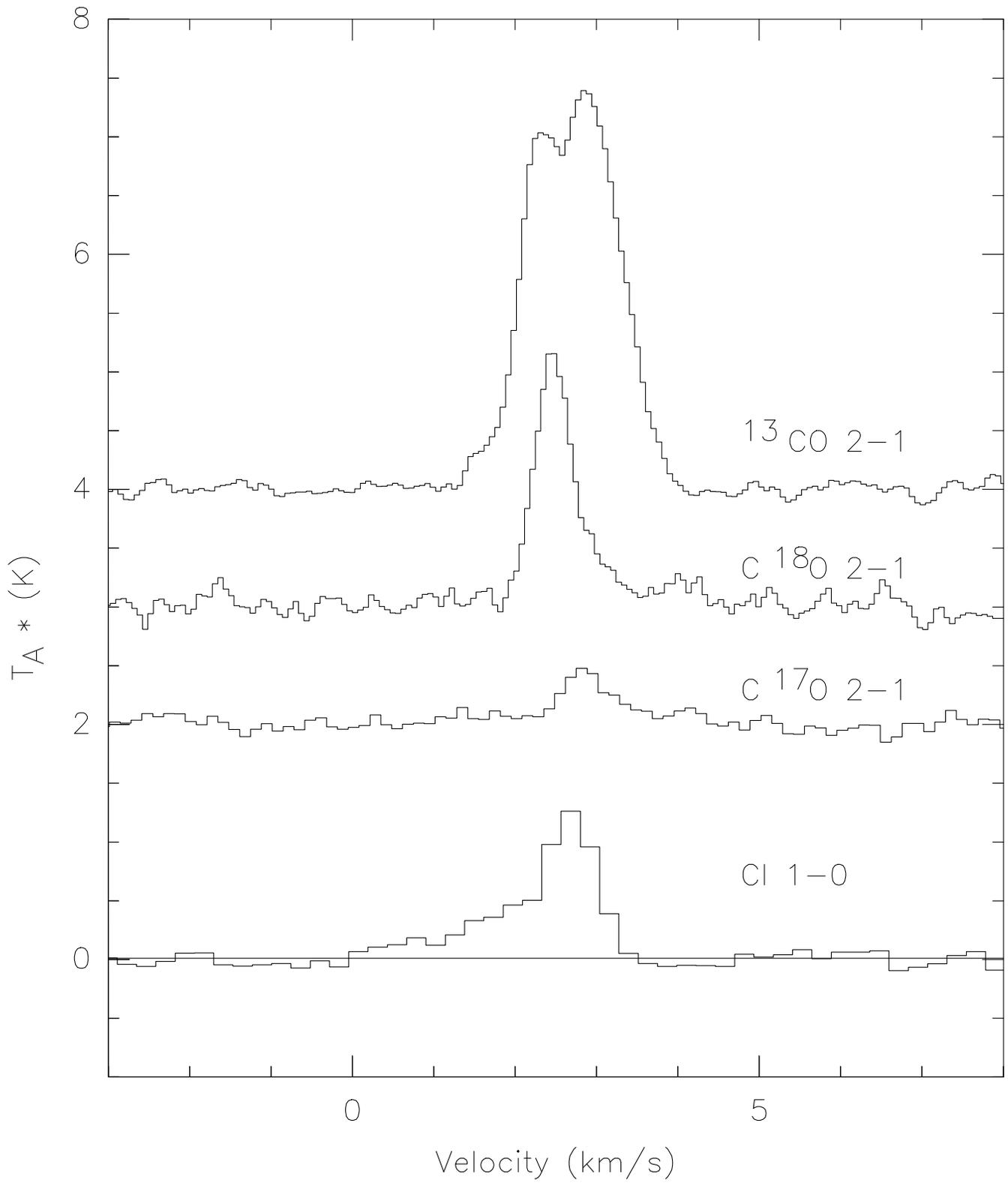
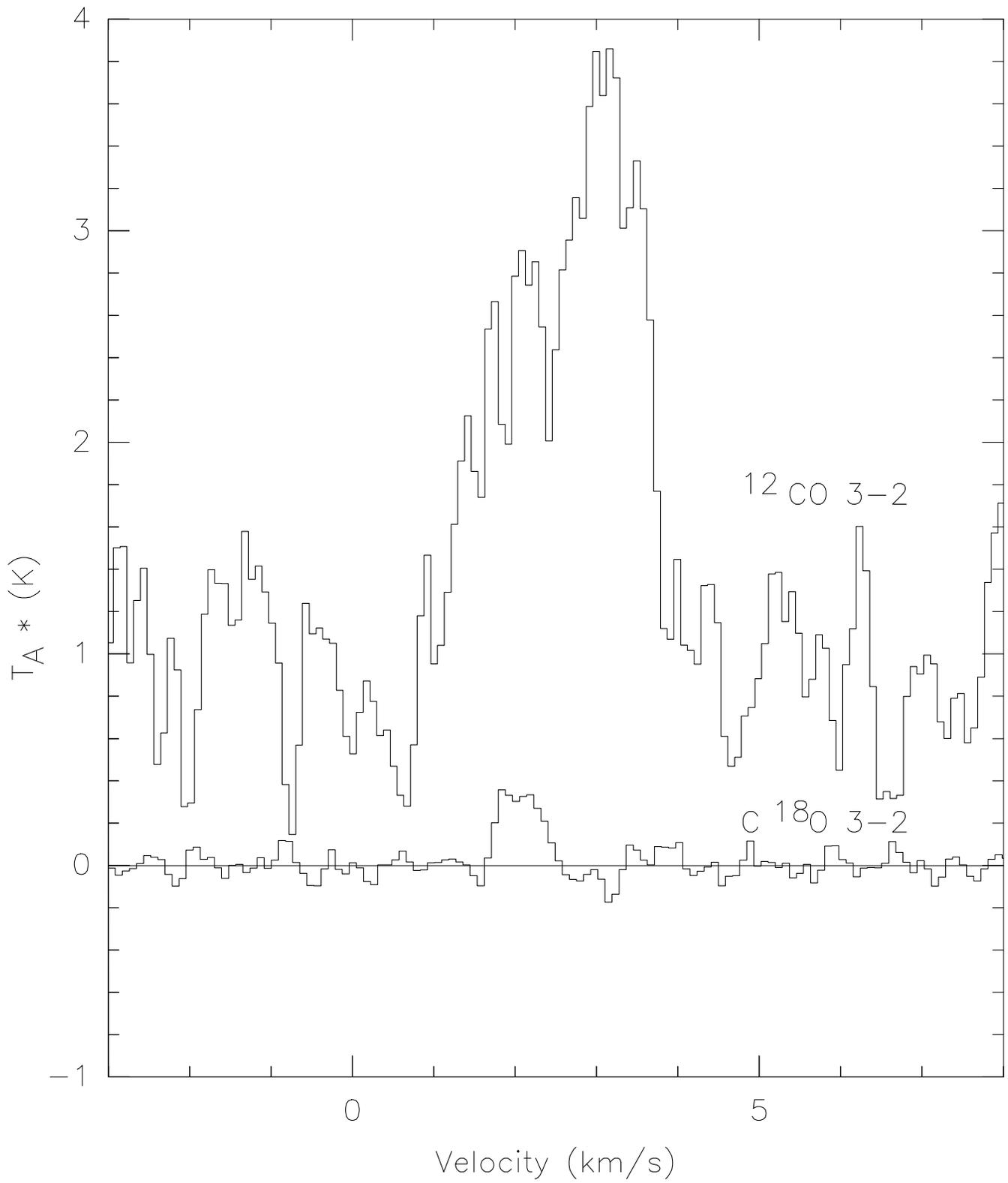
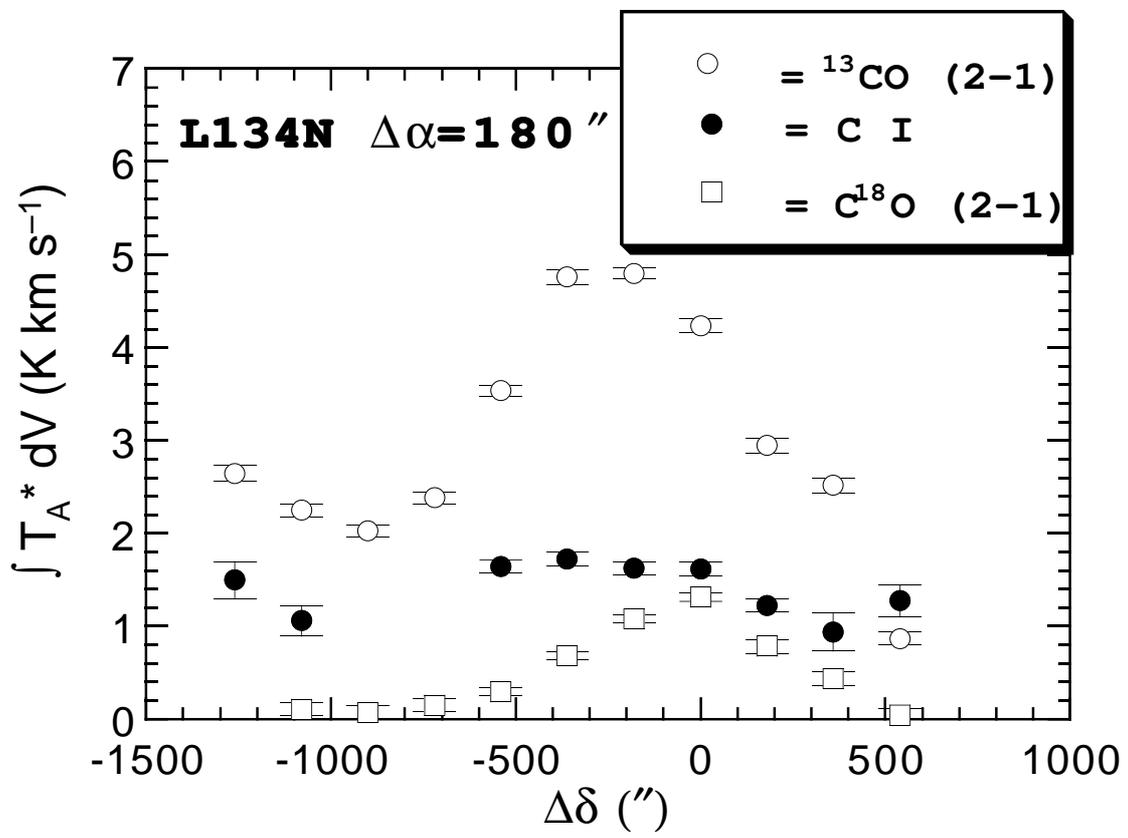
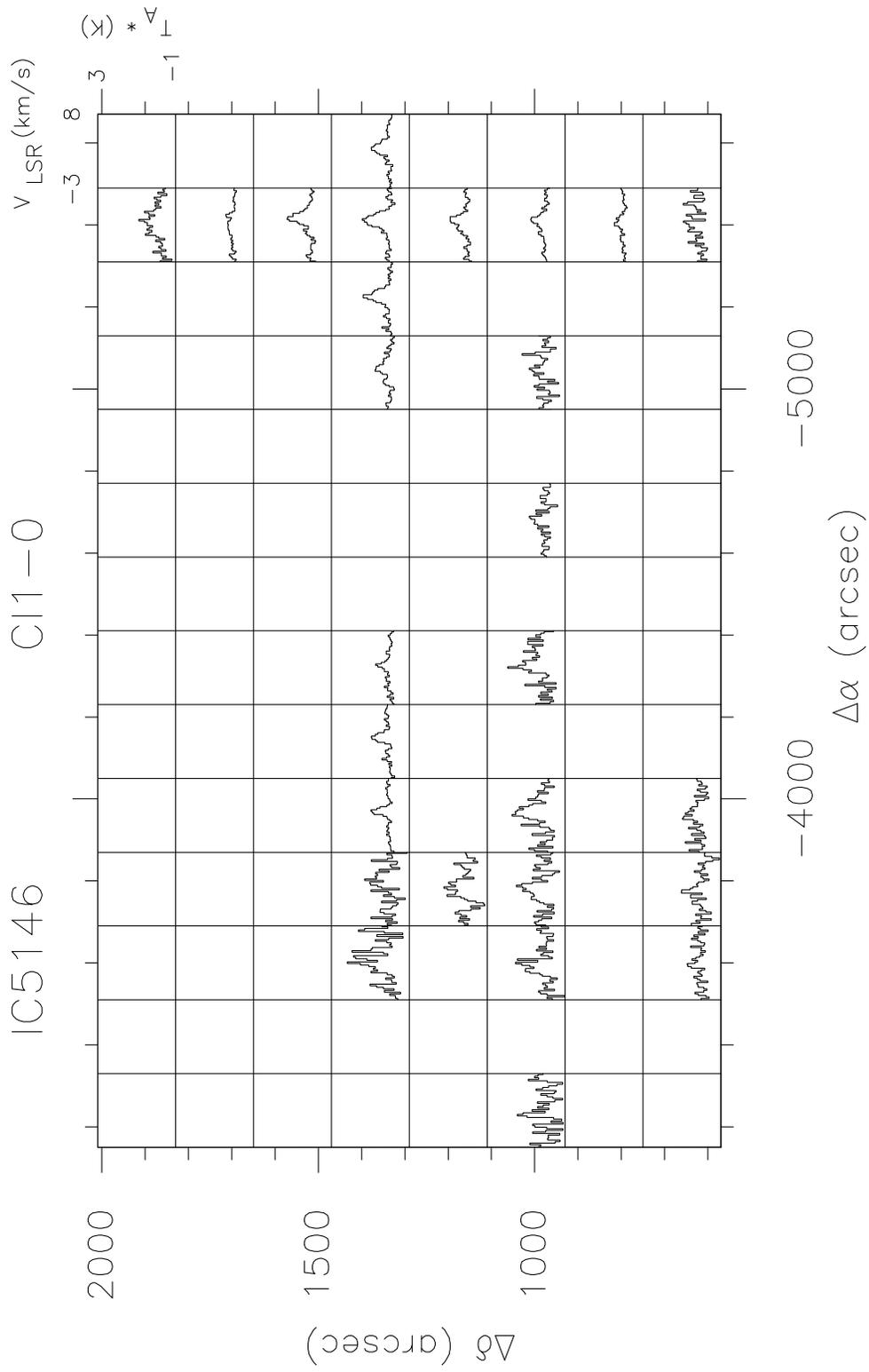
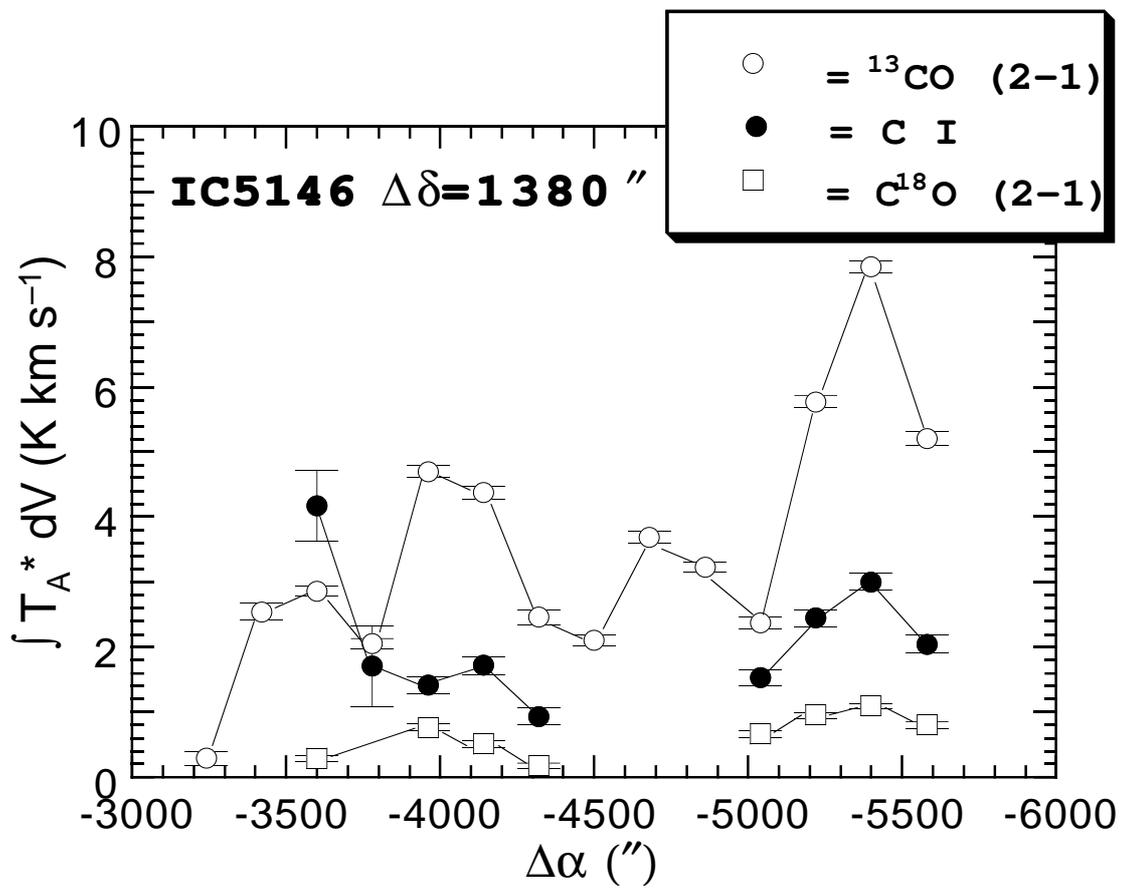
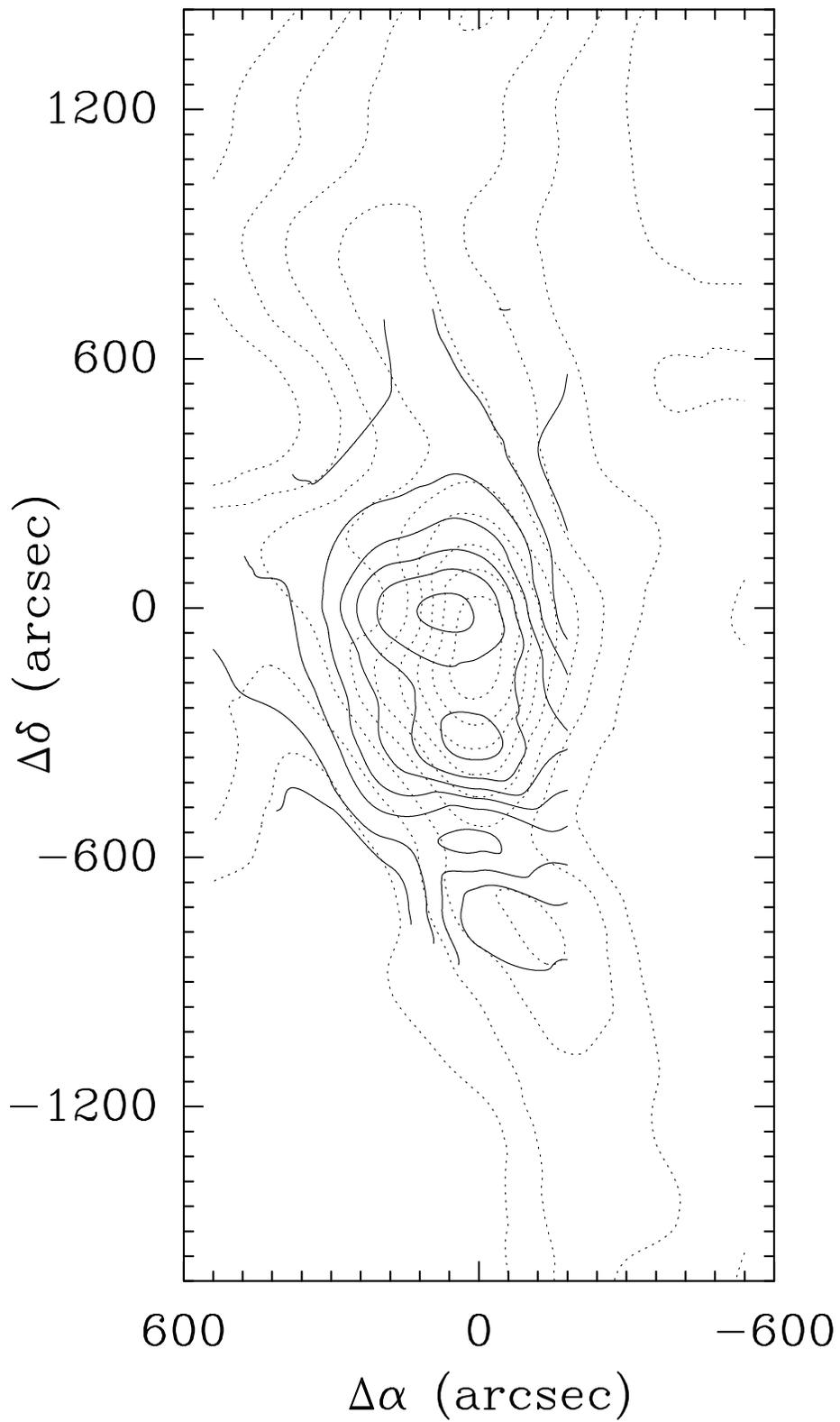
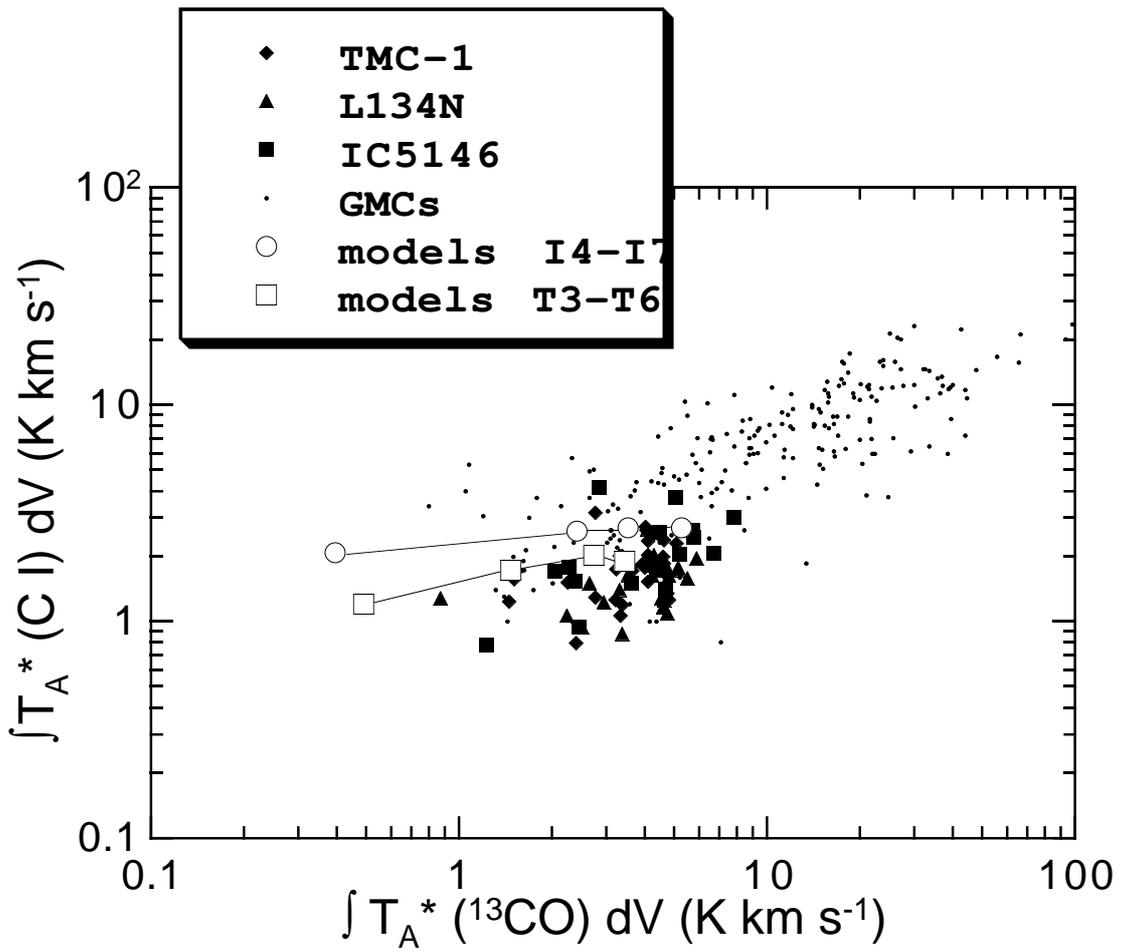
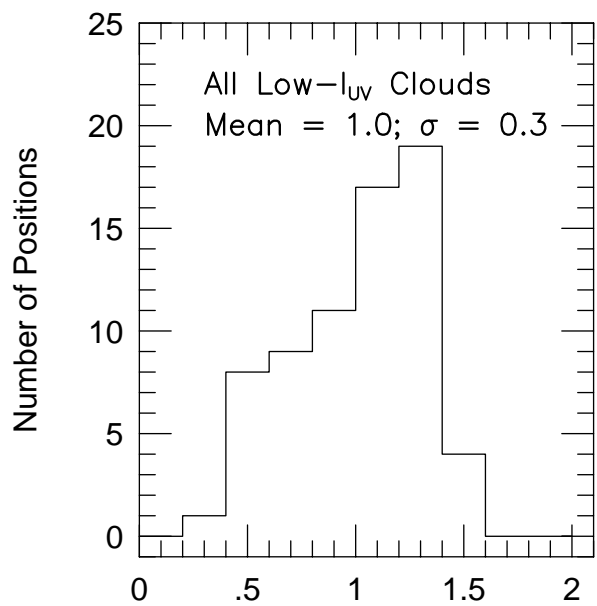
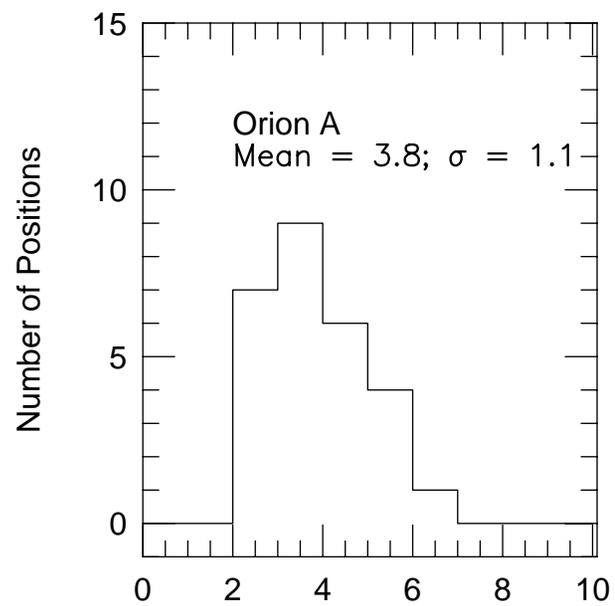
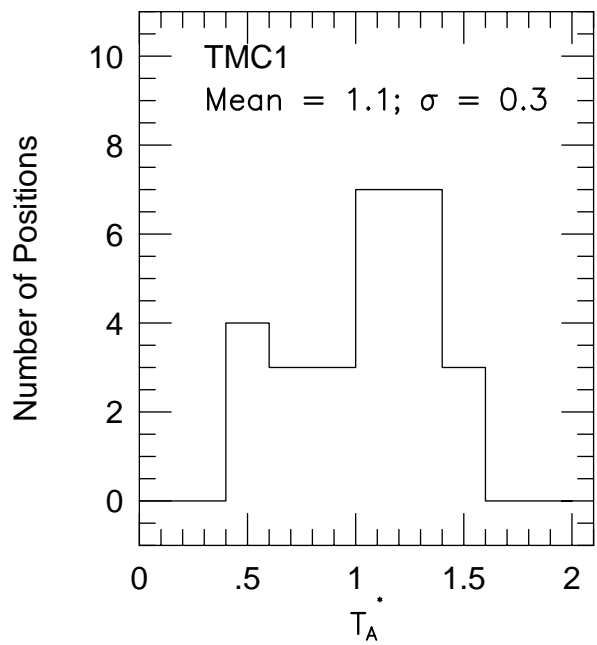
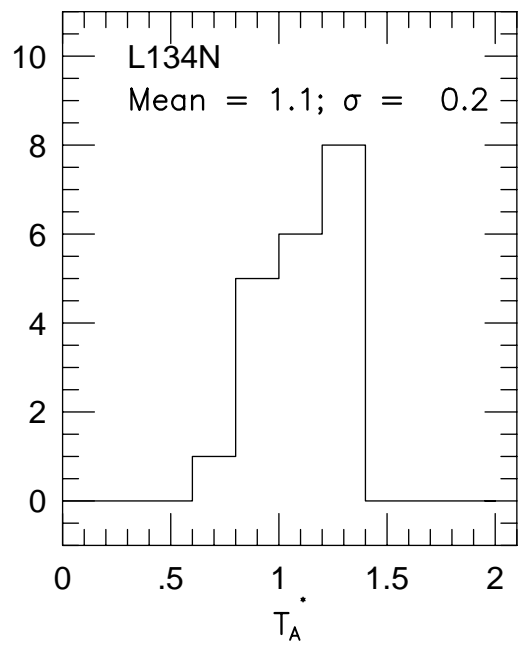
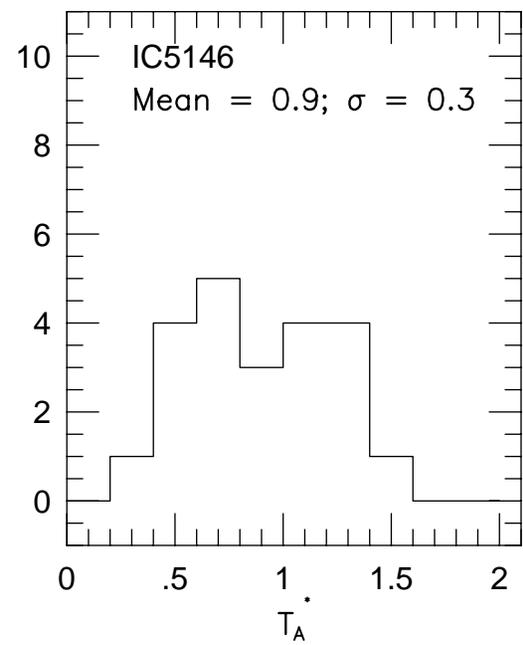
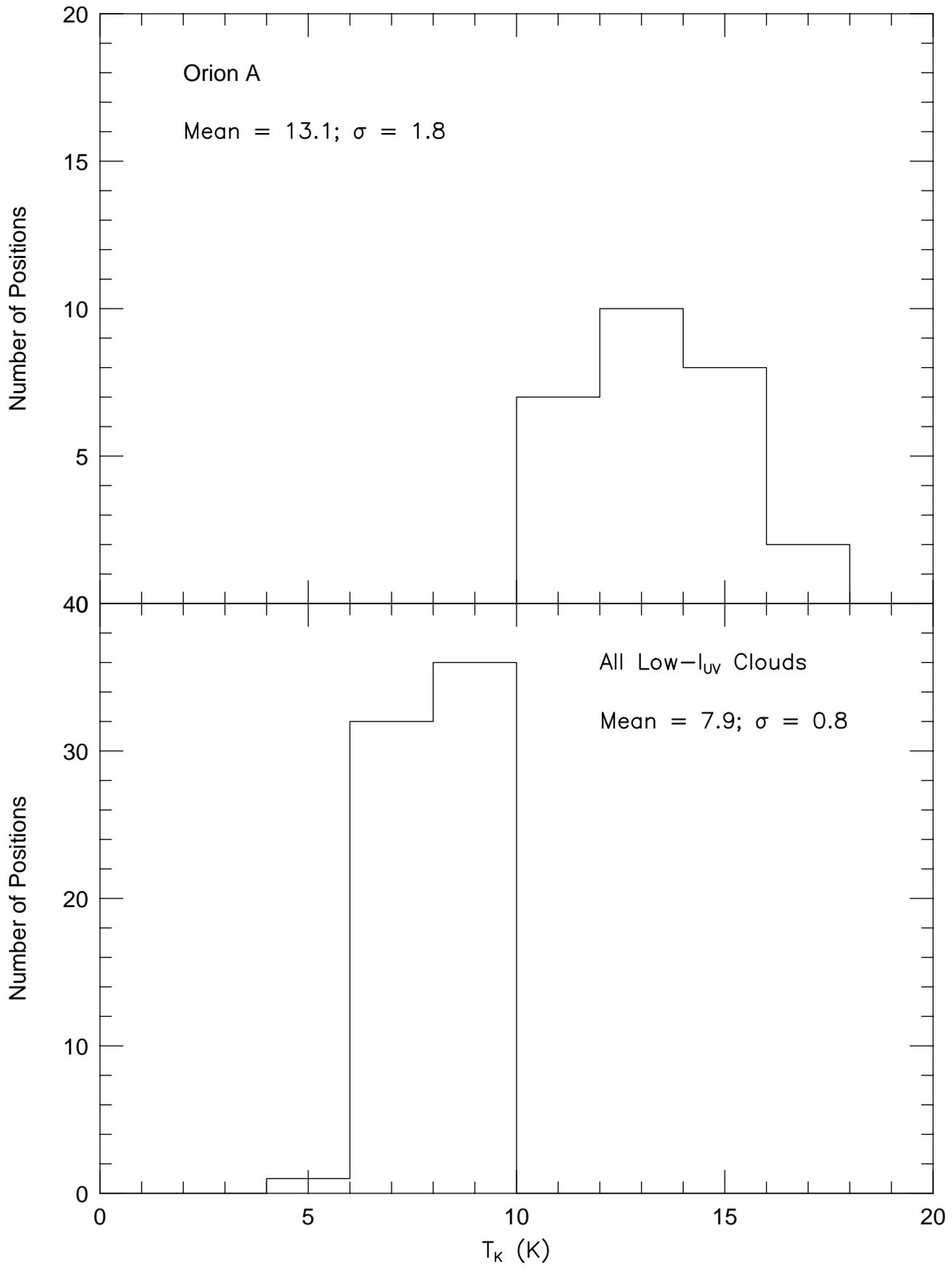
\end{document}